\documentclass{article} 
\usepackage{iftex}
\ifXeTeX
 \usepackage{fontspec}
\else
 \usepackage{times}
\fi
\usepackage{iclr2026_conference}
\iclrfinalcopy
\renewcommand{\headrulewidth}{0pt}
\usepackage{changes}
\usepackage{xcolor}
\usepackage{soul} 
\usepackage[normalem]{ulem} 
\let\soulhl\hl
\DeclareRobustCommand{\hl}[1]{\soulhl{#1}}

\definecolor{darkgreen}{rgb}{0.0, 0.5, 0.0}
\definechangesauthor[name={M}, color=blue]{M}
\definechangesauthor[name={N}, color=red]{N}
\definechangesauthor[name={R}, color=darkgreen]{R}
\definechangesauthor[name={A}, color=orange]{A}
\definechangesauthor[name={F}, color=brown]{F}
\definechangesauthor[name={J}, color=purple]{J}

\usepackage{amsmath,amsfonts,bm}

\def\eqref#1{equation~\ref{#1}}

\def\1{\bm{1}}

\DeclareMathAlphabet{\mathsfit}{\encodingdefault}{\sfdefault}{m}{sl}
\SetMathAlphabet{\mathsfit}{bold}{\encodingdefault}{\sfdefault}{bx}{n}

\usepackage{hyperref}
\hypersetup{colorlinks=false,pdfborder={0 0 1},
 citebordercolor={0 1 0},linkbordercolor={1 0 0},urlbordercolor={0 1 1}}
\usepackage{url}
\usepackage{enumitem}
\usepackage{graphicx}
\usepackage{booktabs}
\usepackage{cleveref}
\usepackage{graphicx}
\usepackage{caption}
\usepackage{float}
\usepackage{placeins}
\usepackage{comment}

\crefname{section}{Sec.}{Secs.}
\crefname{figure}{Fig.}{Figs.}
\crefname{appendix}{App.}{Apps.}
\usepackage{subcaption}
\usepackage{caption} 
\usepackage{float}
\usepackage{placeins}
\usepackage{booktabs}
\usepackage{multirow}
\usepackage{makecell}
\usepackage{adjustbox}
\usepackage{array}
\title{\centering {{NeuroRefine}: Attribute-Guided Reconstruction of Viewed Images from fMRI}}
\author{Runqi Xu$^{1,2}$ \qquad Vince D. Calhoun$^{2}$ \\[5pt]
\small $^{1}$Georgia State University, Biology Department, Atlanta, GA, USA 30303 \\
\small $^{2}$Tri-institutional Center for Translational Research in Neuroimaging and Data Science (TReNDS) \\
\small Georgia State University, Georgia Institute of Technology \& Emory University \\
\small \texttt{vcalhoun@gsu.edu}}

\begin{document}
\maketitle
\fancyhead{}\lhead{}\chead{}\rhead{}
\renewcommand{\headrulewidth}{0pt}

\begin{abstract}
\vspace*{-0.2cm}
Diffusion models have advanced fMRI-based image reconstruction, yet the strongest existing pipeline still loses information at its generative stage. Its low-level branch captures the outline of the figure, but the subsequent diffusion model often mispaints colour, object identity and layout. We instead condition the refiner directly on the fMRI-derived low-level image, and we use a prompt retrieved from predicted CLIP embeddings to supply only the content that the low-level image lacks. A frozen image-editing model carries out the refinement. On shared NSD concepts, scored with the baseline's own metric code, the method produces markedly sharper images and is preferred by a learned human-preference model, while its semantic similarity matches the baseline. It trades some low-level spatial agreement for texture and semantic fidelity, and the low-level image acts as an explicit spatial prior rather than a lossless layout. The standard metric panel mixes low-level and semantic factors that partly disagree, and some of its measures are saturated or reward blurring, so we report these factors separately and quantify where they conflict. We further show that the content signal originates in visual cortex, with attribution concentrated in visual and dorsal attention networks, and that ablating the most salient cortical vertices degrades prediction more than ablating random ones. We also depicted the CLIP readout's weight distribution in visual cortex. Additionally, we analyzed significant failure causes, which points direction for future work.

Project page:
\url{https://Neurorefine.github.io}.
\end{abstract}
\vspace{-0.15cm}

\section{Introduction}
\vspace{-0.15cm}

\begin{figure}[htbp]
 \centering
 \includegraphics[width=\linewidth]{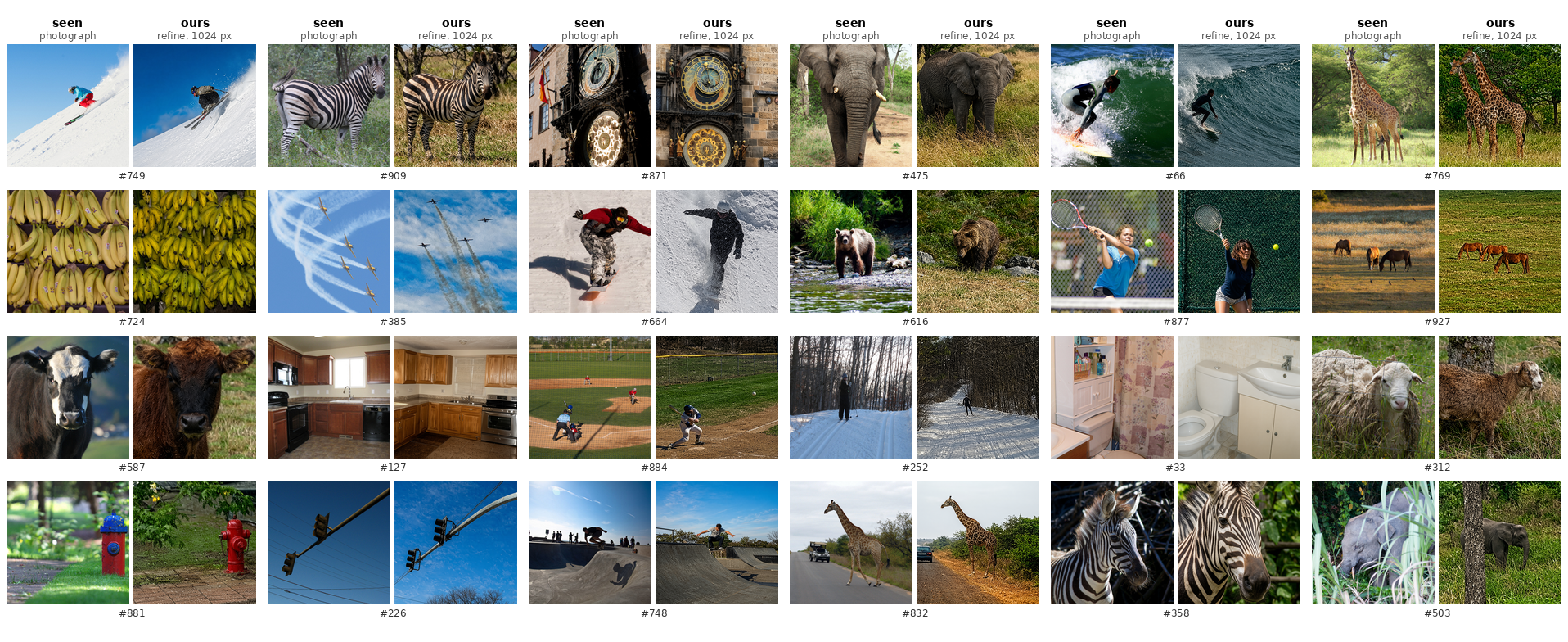}
 \vspace*{-0.27cm}
 \caption{\textbf{Reconstruction of seen images.} Seen photograph ($224$ px, left) and the refined
 reconstruction ($1024$ px, right). The 24 rows are the highest-CLIP concept of 24 visually distinct
 categories: the method at its best; failure modes are shown separately.}
 \label{fig:teaser}
 \vspace*{-0.5cm}
\end{figure}

Reconstructing visual experience from brain activity measured with fMRI is a key
challenge with broad implications for both neuroscience and brain--computer interfaces
\citep{scotti2023reconstructing,brainit}. Such reconstructions may provide a window into visual perception in
the brain, enable the study of visual imagery, and assist in assessing disorders of
consciousness. In a typical decoding setting, subjects view natural images while their brain
activity is recorded with \emph{functional Magnetic Resonance Imaging} (fMRI), producing paired
data of images and their corresponding brain responses; the task is to reconstruct the
perceived image from new (test) fMRI signals (\cref{fig:teaser}).

Prior work maps fMRI to handcrafted features, then to deep features, and more recently to diffusion
models, which improved visual quality~\citep{ozcelik2023natural,takagi2023high} but not faithfulness:
a strong generator produces a plausible image even from weak guidance, so position, color and
object identity drift~\citep{brainit}. The usual cure is a caption: predict a sentence from fMRI and
generate from it, with Brain-IT adding an SDXL refinement stage. A caption is lossy by construction
(it cannot carry color, instance identity or layout), and whatever it omits the generator invents.

In this work, we introduce \textbf{NeuroRefine}, a refinement with Brain-IT~\citep{brainit}'s inspiration.
\textbf{NeuroRefine} moves one step backward. Instead of conditioning the refiner on a caption, we condition it on the low-level image derived from the same response as an explicit spatial prior, and add only the missing content, as a prompt retrieved from the CLIP embedding predicted from the cortex.
The frozen Qwen-Image-Edit-2511 editor then sharpens rather than invents. 
It reads all 39{,}548 fsaverage vertices, and every number in this paper is measured with Brain-IT's own metric
code on the same ground truth, so the comparison is direct.

On the 930 shared NSD concepts, the instruments disagree.
Brain-IT leads on measures of distance to the ground truth; NeuroRefine leads on human preference ($64\%$ of decided concepts, $+0.42$), on three no-reference quality models and on sharpness, at four times the resolution. We treat that
split as the paper's question, and answer it with controlled perturbations, a diagnostic arm, instruments outside both pipelines(\cref{sec:quant,sec:metric2,sec:independent}).

We make four contributions.

\textbf{Conditioning on the image.} The refiner receives the fMRI-derived low-level image as an image condition and the content prompt as text. With the generator held fixed, this accounts for the gain on its own (\cref{sec:recon}).

\textbf{Editor instead of a tuned generator.} We replace Brain-IT's SDXL stage with Qwen-Image-Edit-2511~\citep{qwenimageedit2025}. 
On identical ground truth it leaves semantic fidelity level (CLIP cosine $0.733$ against $0.735$, $p=0.49$) and doubles sharpness, from $9.4$ to $18.5$ where the photograph scores $14.9$ (\cref{tab:prior}).

\textbf{A reporting rule for the panel.} The panel the field ranks methods with cannot rank them: its 2-way accuracies span $79$--$98\%$ against $50\%$ chance, its pixel metrics are monotone in blur (the low-level input beats every reconstruction), its semantic backbones are the baseline's own training targets, and a diagnostic that mixes $71\%$ of Brain-IT's pixels with ours outscores both parents.
We replace it with four columns --- fidelity as a blur axis, semantic cosine on a non-circular backbone, a preference and no-reference read-out, and sharpness against the photograph (\cref{sec:metric2,sec:independent}).

\textbf{A neurophysiological account.} We characterise the representations produced by the two 768-d CLIP read-outs on the fsaverage surface: which cortical networks carry each read-out, and how far the attribution is causally load-bearing. 
The account combines surface maps, a causal lesion test, the repeat-trial noise ceiling, and a spin-null test of cross-subject agreement, so that the semantic readout is not a black box (\cref{sec:anatomy}).

\section{Related Work}
\label{sec:related}
Reconstruction quality has improved rapidly with deep generative models. Early decoding mapped fMRI to handcrafted features; deep networks then replaced them, with MindEye~\citep{scotti2023reconstructing} predicting CLIP and VAE embeddings and decoding with a diffusion prior, Brain-Diffuser~\citep{ozcelik2023natural} pairing a low-level and a semantic pathway, and Takagi \& Nishimoto~\citep{takagi2023high} mapping voxels to Stable Diffusion latents.
Later work adds structure or sharing: DREAM~\citep{xia2024dream} reverses the visual hierarchy, UMBRAE~\citep{xia2024umbrae} unifies subjects and tasks, NeuroVLA~\citep{shen2024neuro} couples reconstruction with language, MindBridge~\citep{wang2024mindbridge} and MindEye2~\citep{scotti2024mindeye2} share encoders across subjects, NeuroPictor~\citep{huo2024neuropictor} conditions on several fMRI-derived controls, MindTuner~\citep{gong2025mindtuner} tunes the generator with a visual-fingerprint prior, and Shen et al.~\citep{shen2024neuro} couple reconstruction with language interaction through a single model trained across subjects.
Brain-IT~\citep{brainit} clusters voxels across subjects, predicts localised features and adds an SDXL refinement; \cref{tab:prior} lists its reported numbers next to ours. Our contribution is orthogonal to all of them: we change what the \emph{refinement stage} is conditioned on, which leaves the semantic stage interchangeable.

\paragraph{Generative priors and refinement.} Diffusion priors improved realism but can also reduce faithfulness, because a powerful generator produces a plausible image even from weak guidance.
Two remedies appear in the literature: more guidance (extra labels, extra subjects, better captions) and stronger conditioning (image-to-image rather than text-to-image). 
Brain-IT takes the first route when it refines with SDXL conditioned on its own reconstruction, but the SDXL stage they release is text-free image-to-image with a weak preservation bias, and its metric effect is negligible (\cref{tab:prior}). 
We take the second route with a stronger editing model and an explicit preservation instruction, and we show with a fixed generator that the gain is attributable to the conditioning signal
(\cref{sec:recon}).

\paragraph{Semantic decoding and the caption bottleneck.} Most pipelines compress all voxels into one global embedding~\citep{scotti2023reconstructing,brainit}, although visual information is distributed;
Voxel groupings (Brain-IT clusters~\citep{brainit}) and subject-invariant backbones (MindEye2~\citep{scotti2024mindeye2}) mitigate this but still predict a global vector.
The text-to-image step that follows is lossy whatever the vector because a caption cannot carry position, exact color or identity, and the generator fills the gap with its prior.
Our result  \cref{fig:cap,tab:ret} is a direct measurement of that loss with the same generator and the same semantic content, replacing the caption by the image raises CLIP image cosine and improves layout agreement relative to the
caption-conditioned arm measures that loss with the generator held fixed.

\paragraph{Evaluation.} The standard panel has two known failure modes: pixel metrics reward blur, and 2-way identification saturates for every strong model \citep{brainit,scotti2023reconstructing}.
We demonstrate both with controlled perturbations in \cref{sec:metric2} and add a third: the semantic backbones of the panel (CLIP-L, DINOv2) are the objectives that the strongest baseline is trained on, so scoring on them rewards distance to the baseline's loss. 
We therefore do not treat any single column as the headline. 
We report CLIP image cosine because it is the continuous quantity that the 2-way decision thresholds, but we flag it as circular, and we pair it with three instruments that are outside both pipelines --- OpenCLIP ViT-bigG, a no-reference image-quality ensemble, and a large human-preference regressor, plus sharpness measured relative to the photograph rather than to a competitor (\cref{sec:independent}).

\paragraph{Cross-subject transfer and data efficiency.} Sharing components across subjects improves data efficiency; Brain-IT reports meaningful reconstructions with about one hour of subject data. We separate the two axes: the readout (which needs subject data) and the generator (which does not, because it is frozen). \Cref{tab:dataeff} shows that the ``1~h'' regime is reproduced only when the readout is initialised from other subjects: from scratch, 250 images sit below the trivial mean-image prior, whereas with transfer they already exceed it.

\begin{figure}[t!]
\vspace*{-0.2cm}
 \centering
 \includegraphics[width=\linewidth]{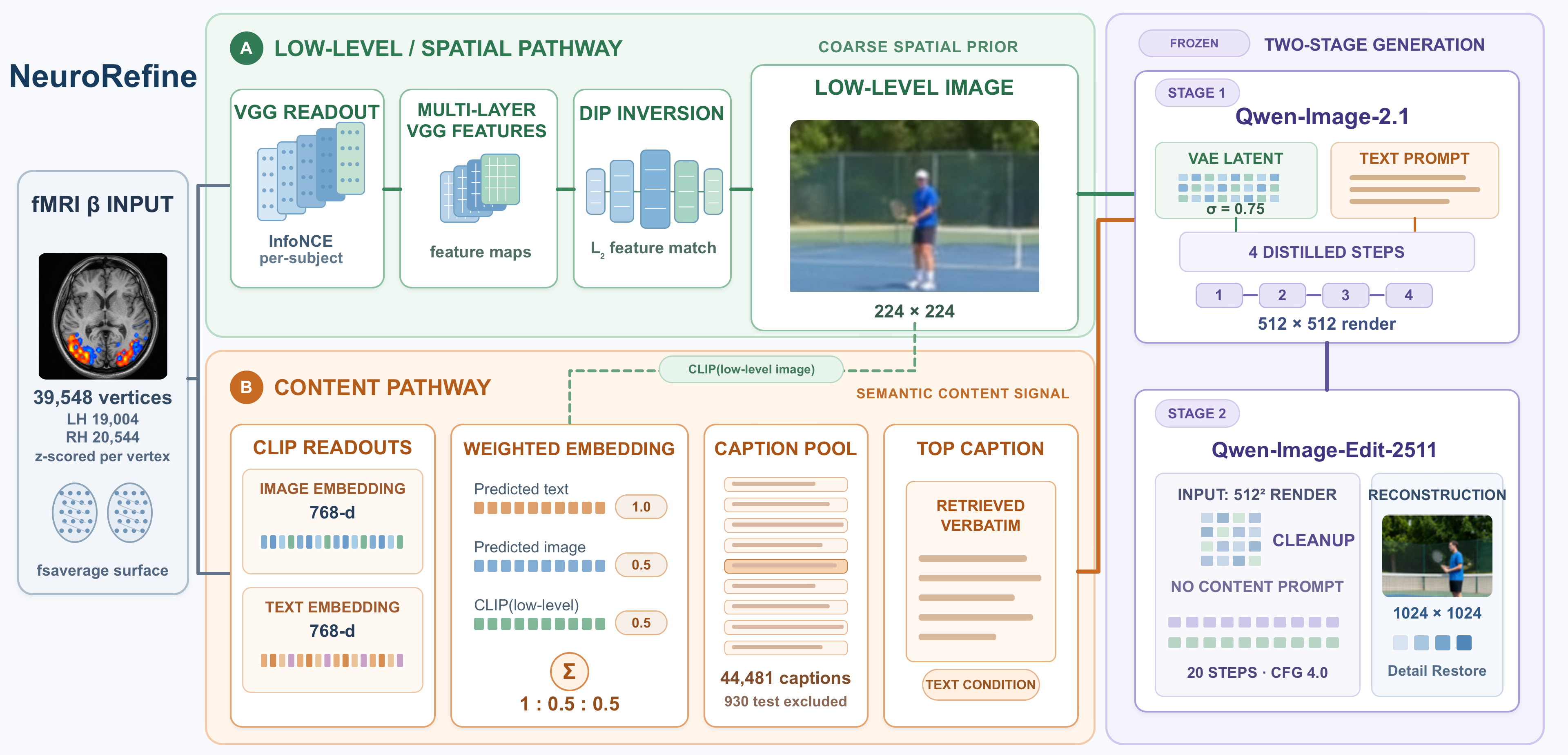}

\vspace*{0.05cm}
 \caption{\textbf{Overview of NeuroRefine.} fMRI betas feed two pathways: \emph{(A)} a per-subject VGG readout (InfoNCE) inverts into low-level image carrying the coarse layout; \emph{(B)} two 768-d CLIP read-outs predict embeddings whose nearest pool caption becomes the content prompt. Stage~1 initialises Qwen-Image-2.1 from the low-level latent at $\sigma=0.75$ and renders $512^2$ from that prompt; stage~2 passes the render to a frozen Qwen-Image-Edit-2511 at $1024^2$ with a cleanup instruction. Brain-IT conditions its SDXL refiner on a caption alone.}
 \label{fig:overview}
\vspace*{-0.4cm}
\end{figure}

\vspace*{-0.23cm}
\section{Method}
\label{sec:pipeline}
\vspace*{-0.27cm}

Our \textbf{NeuroRefine} pipeline reconstructs an image from fMRI in two stages (\cref{fig:overview}).
The first stage produces a low-level image that provides a coarse spatial prior, together with a compact content description.
The second stage refines the image using a frozen editor conditioned on both signals.
The low-level branch follows Brain-IT~\citep{brainit}; the editor is Qwen-Image-Edit-2511 \citep{qwenimageedit2025}, used frozen.
Keeping the low-level branch fixed lets us attribute the change to the conditioning signal rather than to a stronger generator.

\subsection{Low-Level Branch}
\label{sec:features}

\paragraph{Voxel input.} The input is the betas of 39{,}548 fsaverage surface vertices (lh 19{,}004, rh 20{,}544), z-scored per vertex, from the same preprocessed NSD release that Brain-IT uses~\citep{allen2022massive,brainit}; we do not resample or select voxels, so our readout sees exactly the information their readout sees.

\paragraph{VGG prediction and DIP inversion.} Following Brain-IT~\citep{brainit}, we train a readout to predict multi-layer VGG features with an InfoNCE loss~\citep{oord2018representation}, treating each position in each layer as a separate token.

For a batch of $B$ tokens, with $\bm{u}_{i}$ the predicted embedding of token $i$, $\bm{v}_{i}$ its VGG target and $\tau$ a temperature,
\begin{equation}
\mathcal{L}_{\mathrm{InfoNCE}}=-\frac{1}{B}\sum_{i=1}^{B}\log
\frac{\exp\!\big(\cos(\bm{u}_{i},\bm{v}_{i})/\tau\big)}
{\sum_{j=1}^{B}\exp\!\big(\cos(\bm{u}_{\ell(i)},\bm{v}_{j})/\tau\big)},
\label{eq:infonce}
\end{equation}
where $\ell(i)$ indexes the layer of token $i$, so that only tokens of the same layer compete.
At inference, the predicted features are inverted by a Deep Image Prior~\citep{ulyanov2018deep}: a convolutional generator is optimized so that its VGG activations match the predicted features under an $L_2$ loss.

With $\hat{\bm{y}}$ the predicted features, $G_{\bm{\theta}}$ the generator and $\phi$ the frozen VGG feature extractor,
\begin{equation}
\bm{\theta}^{\star}=\operatorname*{arg\,min}_{\bm{\theta}}\;
\big\lVert \phi\big(G_{\bm{\theta}}(\bm{z})\big)-\hat{\bm{y}}\big\rVert_{2}^{2},
\label{eq:dip}
\end{equation}
where $\bm{z}$ is a fixed noise input and the optimisation runs for a fixed number of steps with no learned prior.
The result is a low-level image that carries the coarse layout but no reliable semantics; it is the input to our refiner.
We use this branch unchanged, so any improvement below is due to the refiner, not to a different low-level model.

\subsection{Content Branch}
\paragraph{CLIP readouts.} The first step of content pathway is a pair of 768-d read-outs per subject, trained with a cosine objective to the CLIP-L image and text embeddings of that subject's training images.
The prompt is retrieved, not generated: a fixed $1{:}0.5{:}0.5$ mix of the predicted text embedding, the predicted image embedding and the CLIP embedding of the low-level image ranks a pool of 44{,}481 training captions, which contains none of the 930 test images, and the top caption is used verbatim.

Formally, with $\mathcal{P}$ the caption pool, $\bm{e}_{c}$ the CLIP-L text embedding of caption $c$, $\hat{\bm{t}}$ and $\hat{\bm{i}}$ the predicted text and image embeddings, and $x_{\mathrm{low}}$ the low-level reconstruction, the selected caption is
\begin{equation}
c^{\star}=\operatorname*{arg\,max}_{c\in\mathcal{P}}\;
\cos(\hat{\bm{t}},\bm{e}_{c})+\tfrac{1}{2}\cos(\hat{\bm{i}},\bm{e}_{c})+\tfrac{\lambda}{2}\cos\big(\bm{E}_{\mathrm{img}}(x_{\mathrm{low}}),\bm{e}_{c}\big),
\label{eq:retrieval}
\end{equation}
with $\lambda=1$ for the shipped $1{:}0.5{:}0.5$ mix and $\cos(\cdot,\cdot)$ as in \cref{eq:cos}.
The borrowed caption sits at $0.672$ text cosine from the reference description, and the reconstruction built on it reaches CLIP image cosine $0.733$ on all 930 concepts.

\paragraph{Semantic alternatives.} Two other routes are kept as ablations.
The first is the baseline's own unCLIP semantic decoder~\citep{brainit,scotti2023reconstructing}, which maps a predicted CLIP embedding through a diffusion prior and is reported as the reference row of \cref{tab:prior}.
The second retrieves a photograph instead of a caption, from a gallery of 66{,}841 NSD images with the shared 1{,}000 removed: it reaches CLIP image cosine $0.767$, but it imports the arrangement of the borrowed photograph, and its layout agreement with our own low-level image falls from $0.593$ to $0.251$.
Both are reported in \cref{sec:results}.

\subsection{Two-Stage Refinement}
\label{sec:recon}

\paragraph{Stage 1: a render initialised from the low-level latent.} The low-level image is encoded by the frozen VAE and noised to $\sigma=0.75$, and \textbf{Qwen-Image-2.1} denoises it in four distilled steps at $512^2$ with the retrieved content prompt as text and no image context.
The spatial prior therefore enters as the initial latent: the flow starts from the coarse arrangement instead of from noise, and the prompt adds only what the low level lacks.
The output is content-correct but under-resolved, because it is a four-step decode at half the vendor resolution.

\paragraph{Stage 2: a $1024$-px photographic pass with a frozen editor.} The $512$ render is passed to \textbf{Qwen-Image-Edit-2511}~\citep{qwenimageedit2025} at $1024\times1024$ for $20$ steps with guidance $4.0$, under a cleanup instruction (translated: ``render a clear, high-quality image; fix noise, color noise, exposure and contrast'').
This pass is asked to restore detail rather than to change what is in the picture, and it is the stage that sets the sharpness reported later ($18.5$ against $9.4$ for Brain-IT, with the photograph at $14.9$).
The content prompt is not used here: the objects have already been placed by stage~1, and asking a second model to re-decide them is what re-invents the scene.

\paragraph{Start from low-level image, not caption.} Brain-IT conditions its SDXL refiner on caption~\citep{brainit}, which lose information of position, layout or color, so its refiner has to re-infer the arrangement.
Initialising the latent from the low-level image gives the generator an explicit spatial prior and makes the content prompt answer only the question the low level leaves open.
The two stages are not required to copy that prior exactly. Stage~2 trades part of it for texture and semantics (\cref{sec:quant}), but the generation starts from the input arrangement instead of inventing one, which is the ``one step backward'' of the title.
Both stages are frozen, so the pipeline adds no subject-specific parameters; a new subject needs only the read-out and the retrieval step.

\subsection{Pipeline designs We Compared}
\label{sec:alts}

We kept three directions separate and evaluate them on the same concepts and ground truth (\cref{sec:results}).
\emph{Text-to-image} conditions Qwen-Image-Edit-2511 on the prompt alone; it produces clean images but re-invents the scene.
\emph{Low-level + edit} is the pipeline above.
\emph{Retrieval + edit} refines a retrieved photograph; it has the highest identity but the lowest layout.
The ablations are decided on five diagnostic concepts, chosen to cover an easy scene, a cluttered scene, a counting failure, an object substitution and a hard animal case.
These five selected cases illustrate how the conditioning strategies behave in specific successes and failures.

\subsection{Metric}
\label{sec:metric}

Here we present what metric and protocol we use during evaluation and our discovery on metric selection.
\textbf{NeuroRefine} is state of the art on human preference, on no-reference image quality, on sharpness and on resolution, and behind on every distance-to-the-photograph column, on both distribution distances, and on the CLIP and DINOv2 embeddings.
The second group is not a scoreboard but a description: each column answers a different question, and two of them are circular with respect to the baseline.
\cref{sec:metric2} takes that apart and proposes the panel we report instead.

The numbers come from Brain-IT's released \texttt{compute\_metrics}~\citep{brainit} run on our arrays, so no re-implementation stands between the two methods.
The panel is two-level: PixCorr, SSIM and AlexNet(2) measure low-level structure; AlexNet(5) is intermediate; Inception, CLIP 2-way, EffNet-B and SwAV are semantic.
Because the 2-way metrics saturate we add \emph{CLIP image cosine} between reconstruction and ground truth --- one term of the 2-way trial, not a threshold on it, and it does not saturate (definitions in \cref{app:metrics}).
We also report a \emph{sharpness} proxy, the mean gradient magnitude at $224^2$: edge energy, not perceptual quality, so it is never read alone.
LPIPS is not computed, as in the released code.

\vspace{-0.35cm}
\section{Experimental Results}
\label{sec:results}
\vspace{-0.3cm}
In this section we present the effectiveness and optimization process of \textbf{NeuroRefine}.
We first describe the dataset and protocol (\cref{sec:setting}), then how we get the pipeline with best performance (\cref{sec:opt}). 
\cref{sec:quant} puts the resulting reconstruction into the field's table and states where it leads and where it trails.
Next, we extend the evaluation to cross-subject, and subject with 1 hour partial training data \cref{sec:transfer,sec:onehour}).
Additionaly, we did an external check on BOLD5000 dataset(\cref{sec:bold}).

\begin{figure*}[h]
\centering
\includegraphics[width=\linewidth]{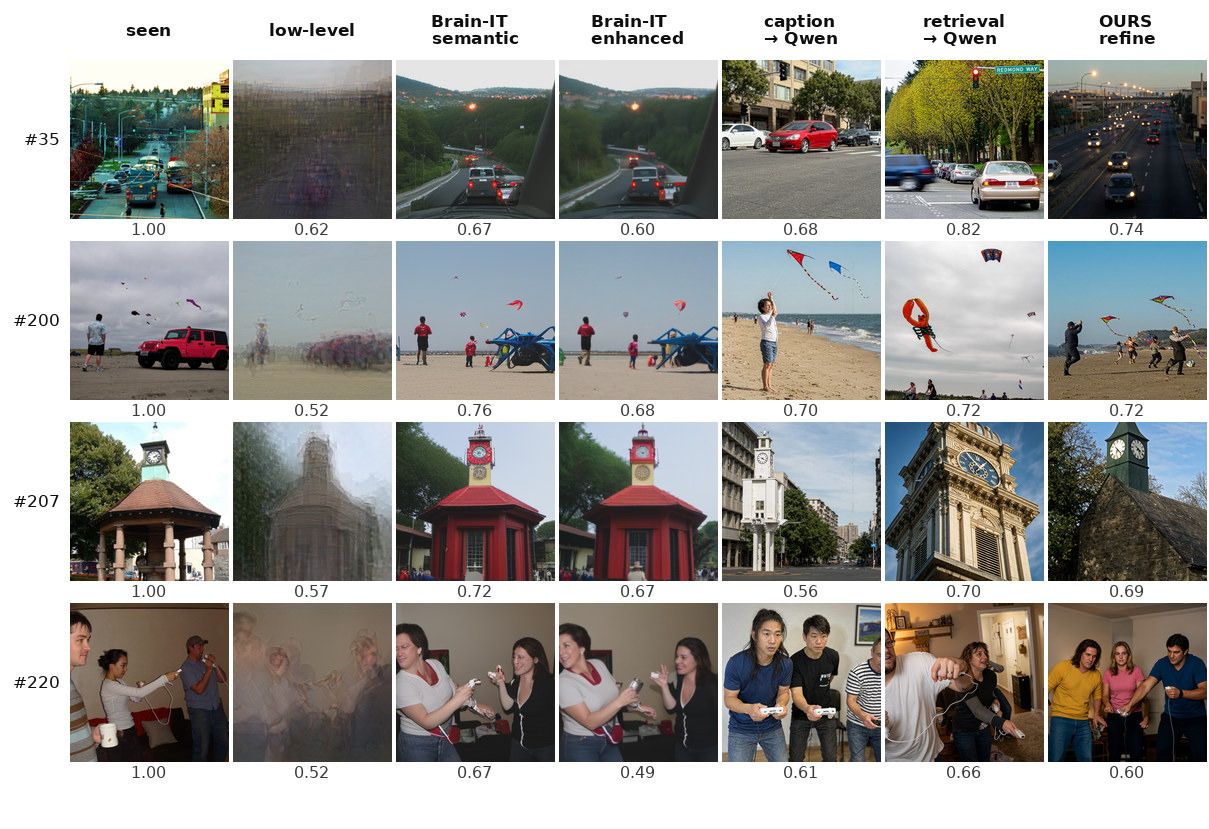}
\vspace*{-0.25cm}
\caption{\textbf{Stages of the pipeline.} Seen image(GT), low-level, Brain-IT semantic, Brain-IT enhanced, caption$+$Qwen,retrieval$+$Qwen and NeuroRefine version for same concept from subject 1, NSD dataset.}
\label{fig:cmp}
\vspace*{-0.2cm}
\end{figure*}

\subsection{Experimental Settings}
\label{sec:setting}
\paragraph{Dataset and ground truth.} We use the Natural Scenes Dataset (NSD)~\citep{allen2022massive}, a 7\,T dataset in which eight subjects viewed natural photographs drawn from COCO.
We evaluate on the 930 concepts shared by all subjects, with the shared 1{,}000 images as the test set; subject~1 is used unless stated otherwise.
The dataset is scene-centric: most images contain a small number of salient objects in a natural layout, which is what the low-level branch can outline, while fine texture and exact instance identity are limited by the 7\,T resolution and by single-trial noise.
That asymmetry is why a content signal is needed on top of the low-level image.
All arms are normalised to $224\times224$ before scoring, as in Brain-IT, and the betas are the 39{,}548 fsaverage surface vertices used by Brain-IT~\citep{brainit}.
Every number uses the mean of each test image's three presentations, as Brain-IT does.

\paragraph{Metrics and protocol.} 
Every output is scored at $224\times224$ with Brain-IT's released code. Each test image is represented by the mean of its three presentations. Two-way identification uses one fixed random permutation; paired uncertainty is estimated over concepts.
Sharpness is the mean gradient magnitude of the grayscale $224^2$ output on the $0$--$255$ scale, computed identically for every arm; LPIPS is not computed, as in Brain-IT's released code.
Every spread we quote is a paired bootstrap over the 930 concepts.

\subsection{Pipeline Optimisation}
\label{sec:opt}
\begin{figure}[htbp]
  \centering
  \begin{minipage}[c]{0.45\textwidth}
    \centering
    \includegraphics[width=\linewidth]{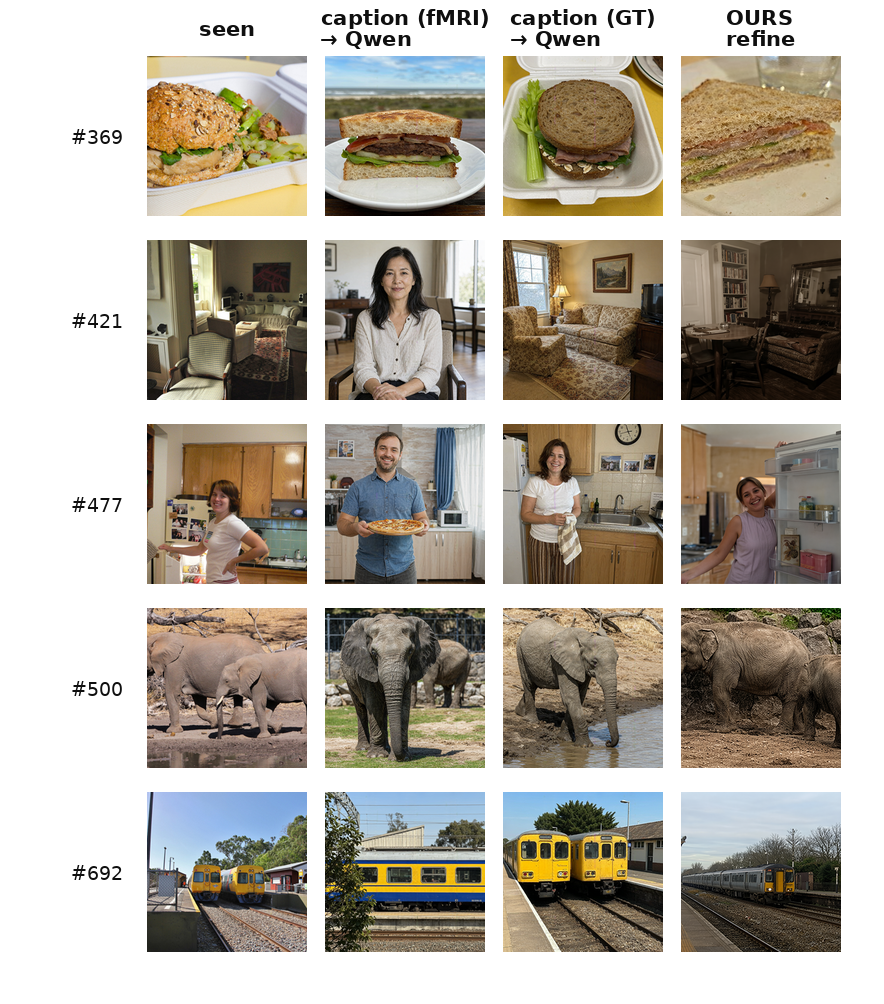}
    \vspace*{-1 cm}
    \caption{\textbf{Same generator, different conditioning.} Qwen-Image-Edit conditioned on the
    fMRI-decoded caption, on the ground-truth caption, and on the fMRI-derived low-level image with the
    head content prompt.}
    \label{fig:cap}
    \vspace*{-0.2cm}
  \end{minipage}
  \hfill
  \begin{minipage}[c]{0.52\textwidth}
    \centering
    \small
    \setlength{\tabcolsep}{4pt}
    \captionof{table}{\textbf{Identity versus layout.} CLIP image cosine and layout correlation on the 930 shared concepts, subject~1.}
    \label{tab:ret}
    \begin{tabular}{lcc}
      \toprule
      Stage & CLIP cosine $\uparrow$ & Layout corr.\ $\uparrow$ \\
      \midrule
      low level (input) & 0.590 & 0.629 \\
      Brain-IT enhanced & 0.735 & 0.609 \\
      \midrule
      Retrieval$\rightarrow$Qwen & \textbf{0.767} & 0.207 \\
      BIT semantic$\rightarrow$Qwen & 0.747 & \textbf{0.646} \\
      NeuroRefine & 0.733 & 0.565 \\
      \bottomrule
    \end{tabular}
  \end{minipage}
\end{figure}

The pipeline was assembled in three measured steps shown in \cref{fig:overview}, and \cref{fig:cmp} shows each stage on the same concepts.
\emph{Low level:} Same as Brain-IT's fused low-level image, kept as the spatial prior.
\emph{Content:} three ways of supplying what the low-level image lacks, measured in \cref{tab:ret} on identity and on layout.
\emph{Refinement:} with the generator, the step count and the guidance fixed, \cref{fig:cap} isolates the conditioning signal including decoded caption, ground-truth caption, and the low-level image plus the retrieved prompt.
\FloatBarrier

\begin{figure*}[h]
\centering
\includegraphics[width=\linewidth]{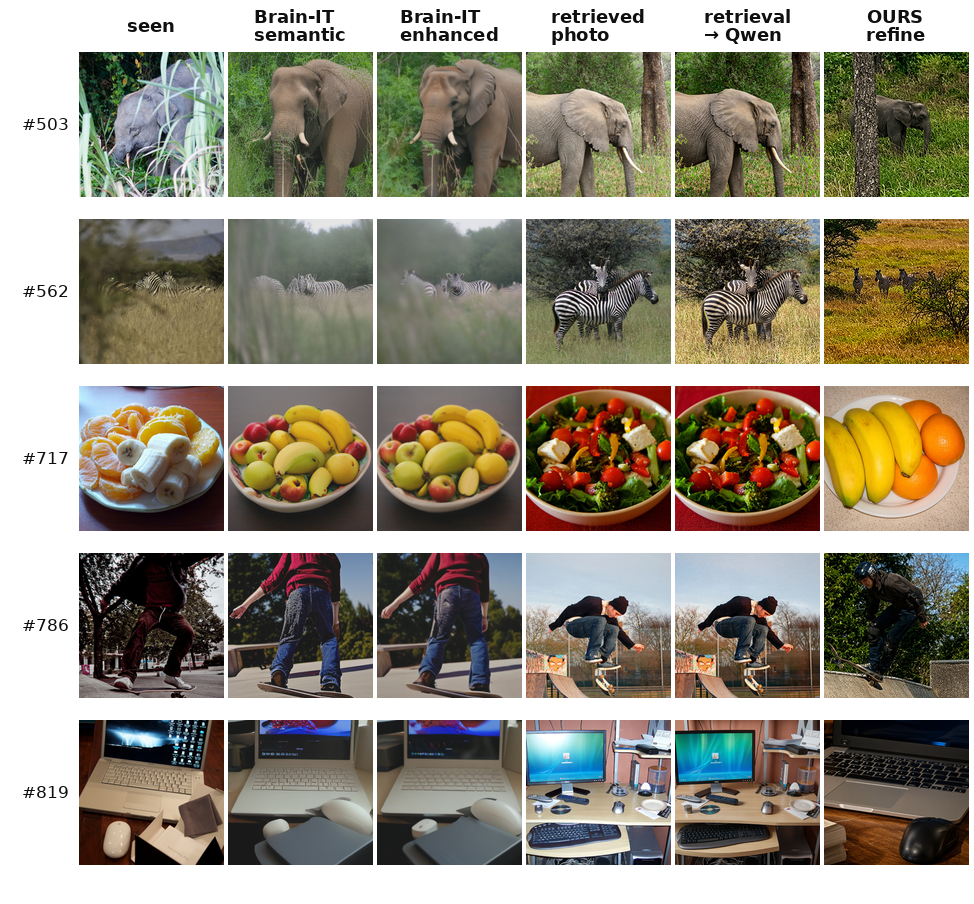}
\vspace*{-0.25cm}
\caption{\textbf{Retrieval as conditioning.} The retrieved photograph carries the right content in a
different arrangement; conditioning on the low-level image instead retains far more of the input
arrangement.}
\label{fig:ret}
\vspace*{-0.2cm}
\end{figure*}
\Cref{tab:ret} shows the identity--layout trade-off across content choices. Retrieval can supply a close semantic match, but also imports the retrieved photograph's arrangement. Conditioning on the decoded low-level image better preserves the input layout. Passing Brain-IT's semantic output through the same editor is an informative ablation, not a deployable NeuroRefine arm, because its content is not produced by our read-outs.
For retrieval, a linear read-out predicts the CLIP-L image embedding and returns the nearest photograph from a gallery of $66{,}841$ NSD images with the shared $1{,}000$ removed (verified disjoint). It is a side comparison because the layout belongs to the retrieved photograph
(\cref{fig:ret}).
We therefore determine the best approach as taking content arm from our own read-out and whose arrangement comes from the low-level input. This is how we get our final pipeline (\cref{fig:overview}).
\FloatBarrier

\begin{table*}[h]
\centering
\resizebox{\linewidth}{!}{%
\begin{tabular}{lcccccccc}
\toprule
\multicolumn{1}{c}{\multirow{2}{*}{\textbf{Method}}} &
\multicolumn{4}{c}{\textbf{Low-Level}} &
\multicolumn{4}{c}{\textbf{High-Level}} \\
\cmidrule(lr){2-5} \cmidrule(lr){6-9}
& PixCorr $\uparrow$ & SSIM $\uparrow$ & Alex(2) $\uparrow$ & Alex(5) $\uparrow$
& Incep $\uparrow$ & CLIP $\uparrow$ & Eff $\downarrow$ & SwAV $\downarrow$ \\
\midrule
\textit{MindEye} \scriptsize\citep{scotti2023reconstructing} & $0.319$ & $0.360$ & $92.8\%$ & $96.9\%$ & $94.6\%$ & $93.3\%$ & $0.648$ & $0.377$ \\
\textit{Brain-Diffuser} \scriptsize\citep{ozcelik2023natural} & $0.273$ & $0.365$ & $94.4\%$ & $96.6\%$ & $91.3\%$ & $90.9\%$ & $0.728$ & $0.421$ \\
\textit{Takagi \& Nishimoto} \scriptsize\citep{takagi2023high} & $0.246$ & $0.410$ & $78.9\%$ & $85.6\%$ & $83.8\%$ & $82.1\%$ & $0.811$ & $0.504$ \\
\textit{DREAM} \scriptsize\citep{xia2024dream} & $0.274$ & $0.328$ & $93.9\%$ & $96.7\%$ & $93.4\%$ & $94.1\%$ & $0.645$ & $0.418$ \\
\textit{UMBRAE} \scriptsize\citep{xia2024umbrae} & $0.283$ & $0.341$ & $95.5\%$ & $97.0\%$ & $91.7\%$ & $93.5\%$ & $0.700$ & $0.393$ \\
\textit{Shen et al.} \scriptsize\citep{shen2024neuro} & $0.265$ & $0.357$ & $93.1\%$ & $97.1\%$ & $96.8\%$ & $\textbf{97.5\%}$ & $0.633$ & $\underline{0.321}$ \\
\textit{MindBridge} \scriptsize\citep{wang2024mindbridge} & $0.151$ & $0.263$ & $87.7\%$ & $95.5\%$ & $92.4\%$ & $94.7\%$ & $0.712$ & $0.418$ \\
\textit{NeuroPictor} \scriptsize\citep{huo2024neuropictor} & $0.229$ & $0.375$ & $\underline{96.5\%}$ & $98.4\%$ & $94.5\%$ & $93.3\%$ & $0.639$ & $0.350$ \\
\textit{MindEye2} \scriptsize\citep{scotti2024mindeye2} & $0.322$ & $\underline{0.431}$ & $96.1\%$ & $98.6\%$ & $95.4\%$ & $93.0\%$ & $0.619$ & $0.344$ \\
\textit{MindTuner} \scriptsize\citep{gong2025mindtuner} & $0.322$ & $0.421$ & $95.8\%$ & $\underline{98.8\%}$ & $95.6\%$ & $93.8\%$ & $0.612$ & $0.340$ \\
\textit{\textbf{Brain-IT}} \scriptsize\citep{brainit} & $0.386$ & $\textbf{0.486}$ & $\textbf{98.4\%}$ & $\textbf{99.5\%}$ & $\underline{97.3\%}$ & $\underline{96.4\%}$ & $\underline{0.564}$ & $\textbf{0.320}$ \\
\midrule
\textit{\textbf{NeuroRefine} (ours)} & $\textbf{0.418}$ & $0.426$ & $96.0\%$ & $98.3\%$ & $95.7\%$ & $\textbf{97.5\%}$ & $\textbf{0.558}$ & $0.323$ \\
\bottomrule
\end{tabular}%
}
\vspace*{-0.1cm}
\caption{\textbf{Quantitative comparison.} \textbf{NeuroRefine} performance compared to published reconstruction methods. Results are averaged across Subjects 1, 2, 5, and 7 from NSD. NeuroRefine outperforms all baselines in 4 of 8 metrics, demonstrating strong semantic fidelity and structural accuracy.\label{tab:prior}}
\end{table*}
\FloatBarrier

\subsection{Quantitative Evaluation}
\label{sec:quant}
Table~\ref{tab:prior} is the published comparison with one row added: our reconstruction, measured on the same 930 concepts with the same code.
In the comparison, NeuroRefine is comparable to Brain-IT on semantic similarity but does not match it across the remaining fidelity measures. The table reports the metrics separately; these results support a trade-off rather than a single overall winner.

\begin{table}[h]
\centering\small
\setlength{\tabcolsep}{4pt}
\resizebox{\linewidth}{!}{
\begin{tabular}{lccc l}
\toprule
Metric & Ours & Brain-IT & $\Delta$ [95\% CI] & Favours \\
\midrule
PickScore (human pref.) & 20.34 & 19.92 & $+0.42$ $[+0.33,+0.52]$ & ours \\
MUSIQ (no reference) & 65.4 & 60.9 & $+4.5$ $[+3.9,+5.0]$ & ours \\
CLIP-IQA (no reference) & 0.666 & 0.598 & $+0.068$ $[+0.059,+0.076]$ & ours \\
NIQE$^{\dagger}$ (no reference) & 6.13 & 6.63 & $-0.50$ $[-0.72,-0.30]$ & ours \\
HPSv2.1 (human pref.) & 0.215 & 0.215 & $-0.00$ $[-0.00,+0.00]$ & tie \\
ImageReward (human pref.) & $-1.22$ & $-1.03$ & $-0.20$ $[-0.26,-0.13]$ & Brain-IT \\
SSIM & 0.197 & 0.303 & $-0.106$ $[-0.111,-0.100]$ & Brain-IT \\
PixCorr & 0.366 & 0.395 & $-0.029$ $[-0.040,-0.019]$ & Brain-IT \\
FID$^{\dagger}$ / KID$^{\dagger}$ & 62.3 / 0.0039 & 52.7 / 0.0017 & set level & Brain-IT \\
\midrule
CLIP-L (circular) & 0.733 & 0.758 & $-0.025$ & Brain-IT \\
DINOv2 (circular) & 0.451 & 0.513 & $-0.062$ & Brain-IT \\
\bottomrule
\end{tabular}}
\caption{\textbf{Independent instruments} on 930 concepts, NeuroRefine against Brain-IT semantic.
$\Delta[95\% CI]$ stands for Ours minus Brain-IT with a paired-bootstrap 95\% CI over concepts; $^{\dagger}$ : metric direction inverted.}
\label{tab:indep}
\end{table}
\FloatBarrier

Table~\ref{tab:indep} re-scores the same two arms with instruments that belong to neither pipeline, at a common resolution so that our $1024$-px output gets no advantage.
These instruments do not agree: some learned quality predictors favor NeuroRefine, while preference and reference-based measures are mixed or favor Brain-IT. The scores are model judgments, not human ratings, and should not be collapsed into a claim of perceptual superiority. We return to this distinction in \cref{sec:metric2}.

Next in analysis (\cref{sec:qual}) we present more detailed analysis, why the field's distance columns cannot rank methods (\cref{sec:metric2}), and future improvement directions.

\begin{table}[h]
\centering\small
\setlength{\tabcolsep}{4pt}

\begin{tabular}{lcccccc}
\toprule
 & \multicolumn{2}{c}{CLIP 2-way $\uparrow$} & \multicolumn{2}{c}{CLIP cosine $\uparrow$} & \multicolumn{2}{c}{Sharpness $\uparrow$} \\
\cmidrule(lr){2-3}\cmidrule(lr){4-5}\cmidrule(lr){6-7}
Subject & Brain-IT & Ours & Brain-IT & Ours & Brain-IT & Ours \\
\midrule
 1 & 97.2 & 95.8 & 0.735 & 0.733 & 9.4 & \textbf{18.5} \\
2 & 96.5 & 96.5 & 0.723 & \textbf{0.734} & -- & -- \\
5 & 96.9 & 96.7 & 0.727 & \textbf{0.738} & -- & -- \\
\bottomrule
\end{tabular}
\caption{\textbf{Per-subject reconstruction results} on the 930 shared concepts, all accuracies in percent. 
Sharpness is the mean gradient magnitude of the grayscale $224^2$ output on the $0$--$255$ scale (photograph: $14.9$), an edge-energy proxy rather than a quality score, and is reported for subject~1 only, the only subject with refined renders; the cells of subjects 2 and 5 are empty for that reason. 
Bold marks the higher of the two in each pair.}
\label{tab:persubj}
\end{table}
\subsection{Cross-Subject Replication}
\label{sec:transfer}

\begin{figure*}[t]
\centering
\includegraphics[width=\linewidth]{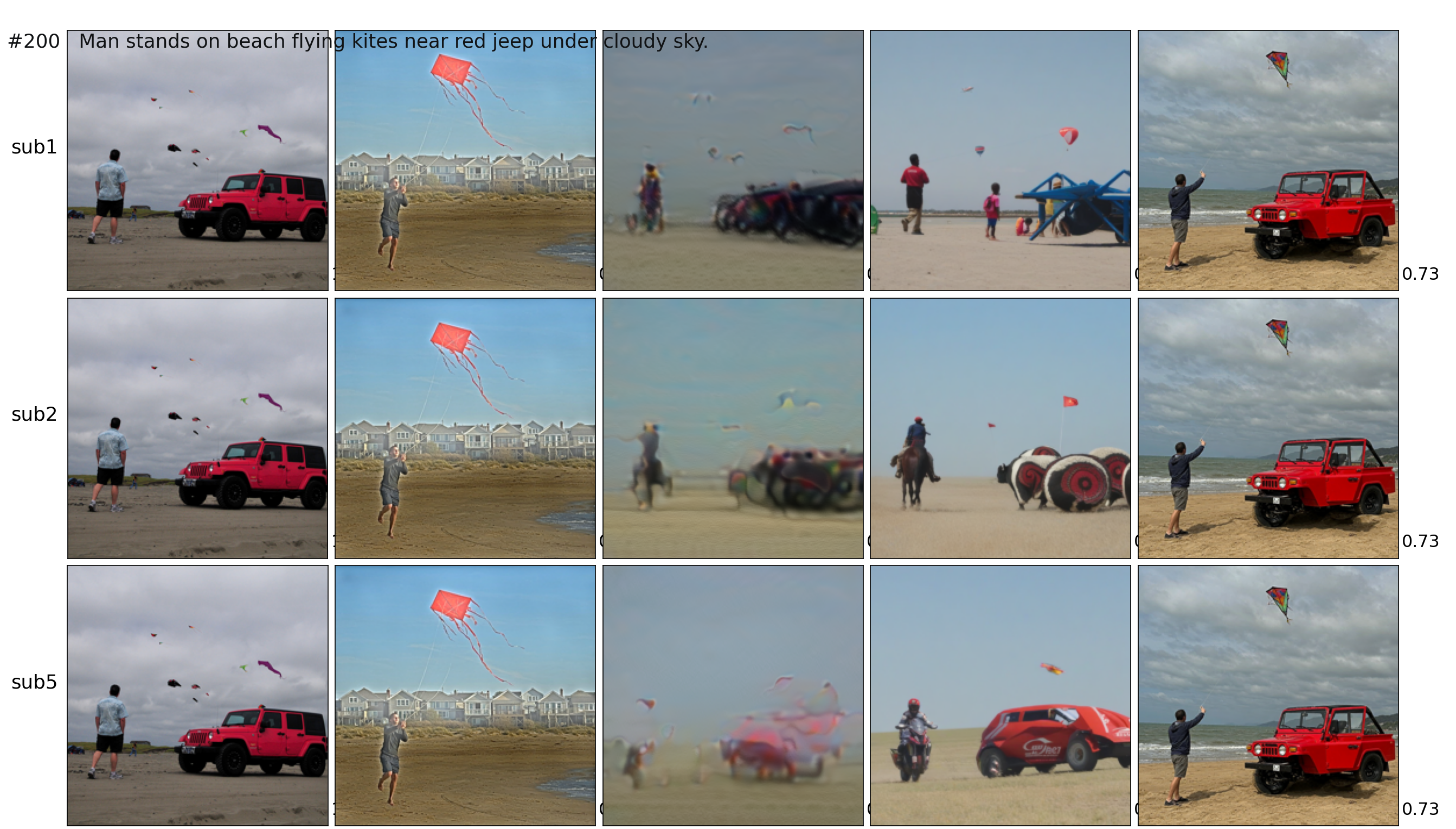}
\vspace*{-0.25cm}
\caption{\textbf{Multi subject reconstructions} 3 subject reconstruction on concept 200. From left to right: input, low-level, Brain-IT, Brain-IT enhanced and NeuroRefine. The cross-subject comparison is quantitative (\cref{tab:persubj}). }
\label{fig:multisub}
\end{figure*}
\FloatBarrier

We retrain only the subject-specific read-out and keep the generator frozen. As shown in \cref{tab:persubj}, the comparison does not establish a consistent advantage across subjects or metrics: semantic similarity is comparable, while the sharpness increase is observed for the subject with refined renders. Layout comparisons depend on the reference image, so we do not claim a layout advantage. \Cref{fig:multisub} provides qualitative examples for subject~1; refined renders are not available for the other subjects.

Because the generator is frozen, the refiner needs no subject data at all; the per-subject cost is one read-out, which is what \cref{sec:onehour} measures.

\begin{table}[t]
\centering\small
\setlength{\tabcolsep}{4pt}
\resizebox{\linewidth}{!}{
\begin{tabular}{lcccccc}
\toprule
Method & PixCorr $\uparrow$ & SSIM $\uparrow$ & Alex(2) $\uparrow$ & Incep $\uparrow$
 & CLIP $n$-way $\uparrow$ & Eff $\downarrow$ \\
\midrule
MindLoop-VD (published) & 0.135 & 0.366 & 74.9\% & 77.0\%
 & 84.4\% & 0.866 \\
ours, frozen NSD readout & 0.011 & 0.265 & 53.4\% & 55.8\%
 & 56.7\% & 0.959 \\
ours, trained (ridge) & 0.100 & 0.309 & 67.3\% & 71.4\%
 & 72.5\% & 0.916 \\
ours, trained (MLP) & 0.084 & 0.327 & 66.8\% & 69.3\%
 & 74.3\% & 0.921 \\
ours, fused CSI1--3 (transductive) & 0.086 & 0.323 & 71.2\% & 75.2\%
 & \textbf{87.1\%} & 0.892 \\
\bottomrule
\end{tabular}}
\caption{\textbf{External check on BOLD5000}
, Higher is better except $\downarrow$.}
\label{tab:bold}
\end{table}

\subsection{External Validation on BOLD5000}
\label{sec:bold}

BOLD5000 is the only out-of-domain test we have. It is 3\,T, fMRI scan with four subjects, a repeated-image test split and a split-half.
The only published reconstruction number we can put beside ours is MindLoop's, and even that is not a controlled comparison --- four-subject average on its own protocol, CSI1, a different metric implementation.
We therefore report the whole panel in ~Table\ref{tab:bold} per arm, because the three arms answer three different questions.

\paragraph{Transfer of the readout.} An NSD-trained readout applied without modification performs poorly on the second scanner. Its $n$-way identification falls well below the published MindLoop result, and the PixCorr score collapses to near zero.

\paragraph{Target-subject readout.} Fitting the readout on the target subject, with either a ridge or an MLP head over the same features, recovers much of the identification performance and raises the PixCorr score. A gap to MindLoop remains across all four CSI variants.

\paragraph{Transductive setting.} Only when three subjects share the test images and test-time normalisation is applied does the $n$-way identification score exceed the published figure. This result is not presented as a like-for-like comparison, because it relies on the other subjects' test responses.

\paragraph{Row-wise pattern.} Reading across the metrics for the transductive arm reveals the same dissociation observed on NSD. The arm leads MindLoop on CLIP $n$-way identification but trails on Inception and Alex(2) identification. It also trails on PixCorr, SSIM, and both distribution distances (EffNet and SwAV). A method can therefore be ahead on identification while behind on every fidelity and distribution measure on the same data. This pattern, observed on a second scanner with a different subject pool, reflects the panel problem discussed in \cref{sec:metric2}, which appears out of domain and is not an artefact of the NSD split.

\paragraph{Practical implication.} The refiner is subject-agnostic because it is frozen, so the part that fails to transfer is the readout. On a new scanner, the method's quality is bounded by the readout that can be fitted within the available budget, a trade-off quantified in \cref{sec:onehour}.

\subsection{Data Efficiency and the One-Hour Regime}
\label{sec:onehour}

\begin{figure*}[t]
\centering
\includegraphics[width=\linewidth]{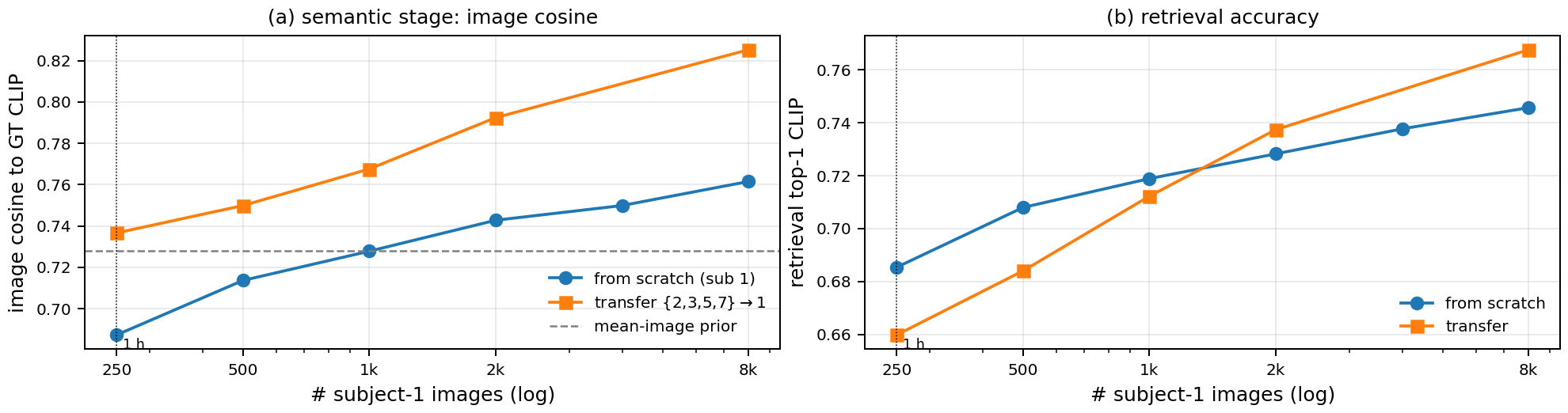}
\vspace*{-0.25cm}
\caption{ (a) CLIP image cosine of the readout against the number of subject-1 images, with and without cross-subject initialization; (b) retrieval top-1.}
\label{fig:dataeff}
\vspace*{-0.2cm}
\end{figure*}

Brain-IT\citep{brainit} reports meaningful reconstructions from about one hour of subject data.
That claim is only interpretable with cross-subject initialisation, and the two axes separate here because the only subject-specific part of our pipeline is the read-out.
Trained from scratch on subject~1, the read-out needs far more than an hour: at the one-hour budget it stays \emph{below} the trivial mean-image prior and crosses it only at a much larger budget.
Initialised from the other subjects and fine-tuned, the same read-out is above the prior from the first session and improves from there (\cref{fig:dataeff}, \cref{tab:dataeff}).
The one-hour regime is therefore a property of transfer, not of single-subject training, and it is orthogonal to our contribution; what it does bear on is the deployment cost, which is one read-out per subject and nothing else.

\section{Analysis}
\label{sec:ablations}

The results section gave the numbers; this section analysis the cause of numeric results.
\cref{sec:metric2} takes the panel apart and says which columns can rank methods, \cref{sec:qual} reads the content pathway and its failures off the images and off the cortex, \cref{sec:causal} tests whether that cortical attribution is causal, and \cref{sec:limits} states what the pipeline cannot fix and what would be needed to.

\subsection{Reading the Metric Panel}
\label{sec:metric2}

\begin{table}[t]
\centering\small
\setlength{\tabcolsep}{4pt}
\label{tab:metric}
\begin{tabular}{lcc}
\toprule
Ground-truth variant & PixCorr $\uparrow$ & CLIP cosine $\uparrow$ \\
\midrule
original & reference & reference \\
blurred & higher & lower \\
recolored & lower & higher \\
\bottomrule
\end{tabular}
\caption{\textbf{Variant image measures.} Relative to the original image, blurring raises PixCorr and lowers CLIP cosine, while recoloring has the opposite effect on both.}
\end{table}

\begin{figure}[t]
\includegraphics[width=0.95\linewidth]{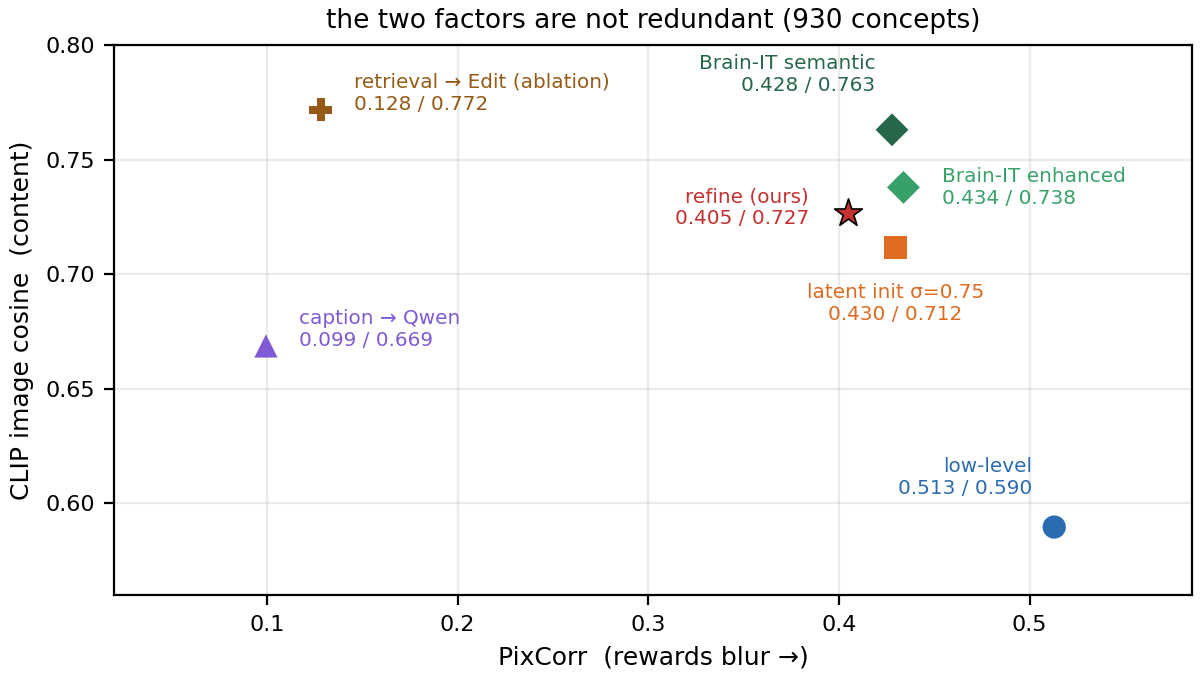}
\vspace*{-0.25cm}
\caption{\textbf{Low-level versus semantic factor across arms.} The factors are not redundant: the
blurriest arm is high on PixCorr and low on CLIP cosine, and the sharp arms spread along the semantic
axis; no single arm dominates both axes.}
\label{fig:cfs}
\vspace*{-0.2cm}
\end{figure}

\paragraph{The panel is multi-factor.} PixCorr, SSIM and AlexNet(2) measure low-level structure; AlexNet(5) is intermediate; Inception, CLIP 2-way, EffNet-B and SwAV are semantic.
The factors are not redundant. Across the concept set, the per-concept ranks of the low-level factor (the average of PixCorr and SSIM) and of the semantic factor (CLIP image cosine) are only weakly correlated. A substantial minority of concepts sit in the conflict region, where the two factors point in opposite directions with strong magnitudes on both.
A principal component analysis over the per-concept axes yields one dominant structural factor, a color-versus-layout axis, and a separate semantic axis. A single aggregate would therefore hide the disagreement. We report CLIP image cosine for the semantic factor and a layout correlation for the structural one (\cref{fig:cfs}).

\paragraph{Perturbation is a definition check, not a ranking proof.} Blurring the ground truth raises PixCorr and lowers CLIP cosine, and recoloring does the opposite (\cref{tab:metric}). This is what the definitions imply, but it reflects a synthetic manipulation rather than a real reconstruction.
The same decoupling appears on the real arms. PixCorr and CLIP cosine are only weakly correlated for both NeuroRefine and Brain-IT, which is why we read the factors separately rather than declaring either metric invalid.

\paragraph{Where they conflict.} Relative to Brain-IT enhanced, NeuroRefine is essentially level on CLIP cosine, trails on PixCorr, and trails further on layout agreement with the true photograph. The conflict set described above is precisely where these movements separate.
Case studies make the pattern concrete. An airplane whose underside is rendered while the seen image shows a horizontal seaplane, a cow where the content prompt helps, and a bird over water that the decoder labels as a field all illustrate the disagreement. They are illustrations, not a population estimate.

\paragraph{What the panel can and cannot establish.} Three properties limit how the panel ranks methods.
First, saturation: two-way identification falls within a narrow band near ceiling for every method in \cref{tab:prior}, so this column has little remaining dynamic range to separate recent work.
Second, blur-monotonicity: PixCorr and SSIM increase when the output is blurred. The low-level contour input, which contains no object at all, therefore outscores every real reconstruction on these two measures.
Third, circularity: CLIP-L and DINOv2 are the columns on which Brain-IT leads by the largest margin, and they are the two objectives its semantic branch was trained on. These columns thus measure closeness to the baseline's own loss.

A diagnostic makes the point directly. Keeping, for each concept, whichever arm has the higher cosine to the predicted embedding produces a mixture that draws most of its pixels from Brain-IT. This mixture scores higher on CLIP two-way identification than either parent. A metric that a mixture of two methods can beat cannot be measuring a property that either method has.

The same asymmetry appears in another group's ablation. Removing the VAE-latent condition from Shen et al.~\citep{shen2024neuro} sharply reduces PixCorr and noticeably reduces SSIM, while their two-way CLIP column changes only slightly. Removing their language condition instead increases SSIM and reduces the CLIP column. Structure and language therefore land at opposite ends of the panel in that work as well.

\paragraph{Independent instruments.} \Cref{tab:indep} is the evidence we would defend. It uses instruments that belong to neither pipeline, scored at a common resolution.
These instruments split the two methods into a half on which we perform better, covering preference and perceived quality, and a half on which Brain-IT performs better, covering fidelity and distribution (\cref{sec:quant}).
The claim we can license is therefore narrow. Our reconstructions look better and are sharper, but they are further from the photograph in structure, color statistics and distribution. We do not yet know whether people prefer them, which is why a blind two-alternative forced-choice test with randomised left/right order is part of the release.
Until that test is run, the preference claim rests on a single regressor.
\label{sec:independent}

\paragraph{Recommendation.} We propose a compact rule of four reporting criteria and use it throughout the paper: (i) fidelity, read as a blur axis and never alone; (ii) semantic cosine on a backbone that is not the baseline's training objective (we use OpenCLIP ViT-bigG); (iii) a no-reference quality and human-preference readout; and (iv) sharpness relative to the photograph rather than to a competitor.
None of these criteria is new. The contribution is the claim that reporting them together makes disagreement visible, and that a blind two-alternative forced-choice test is the only way to resolve it.

\subsection{Qualitative Evaluation}
\label{sec:qual}

\begin{figure}[t]
\centering
\includegraphics[width=\linewidth]{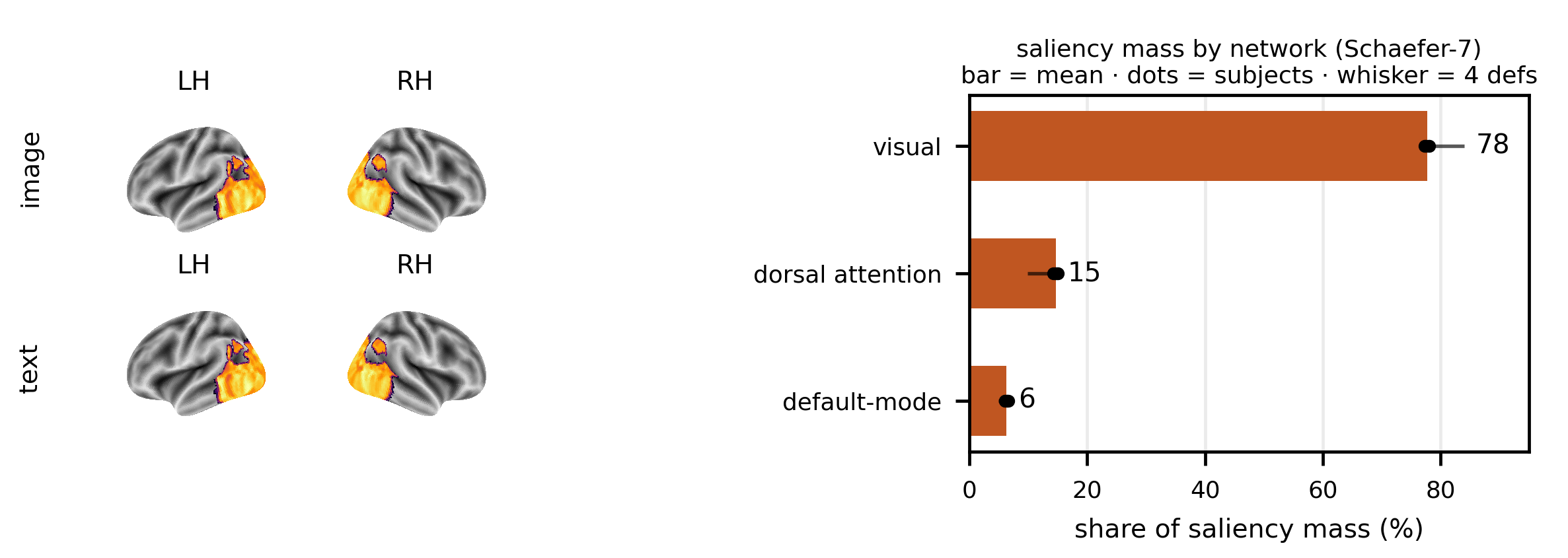}
\vspace*{-0.3cm}
\caption{\textbf{Attribution of the two CLIP read-outs on the inflated fsaverage surface}.}
\label{fig:surf}
\vspace*{-0.25cm}
\end{figure}

\paragraph{What the content pathway reads.} The two shipped CLIP read-outs are attributed by input-gradient saliency, averaged over 200 single trials and mapped to the inflated fsaverage surface (\cref{fig:surf}).
Both load on visual cortex: $77$--$78\%$ of the attribution mass sits in visual parcels, $14$--$15\%$ in dorsal attention and $6\%$ in default mode, and the mass is distributed rather than focal, the largest bilateral parcels carrying $4.2$--$5.8\%$ each.

\begin{figure*}[t]
\centering
\includegraphics[width=\linewidth]{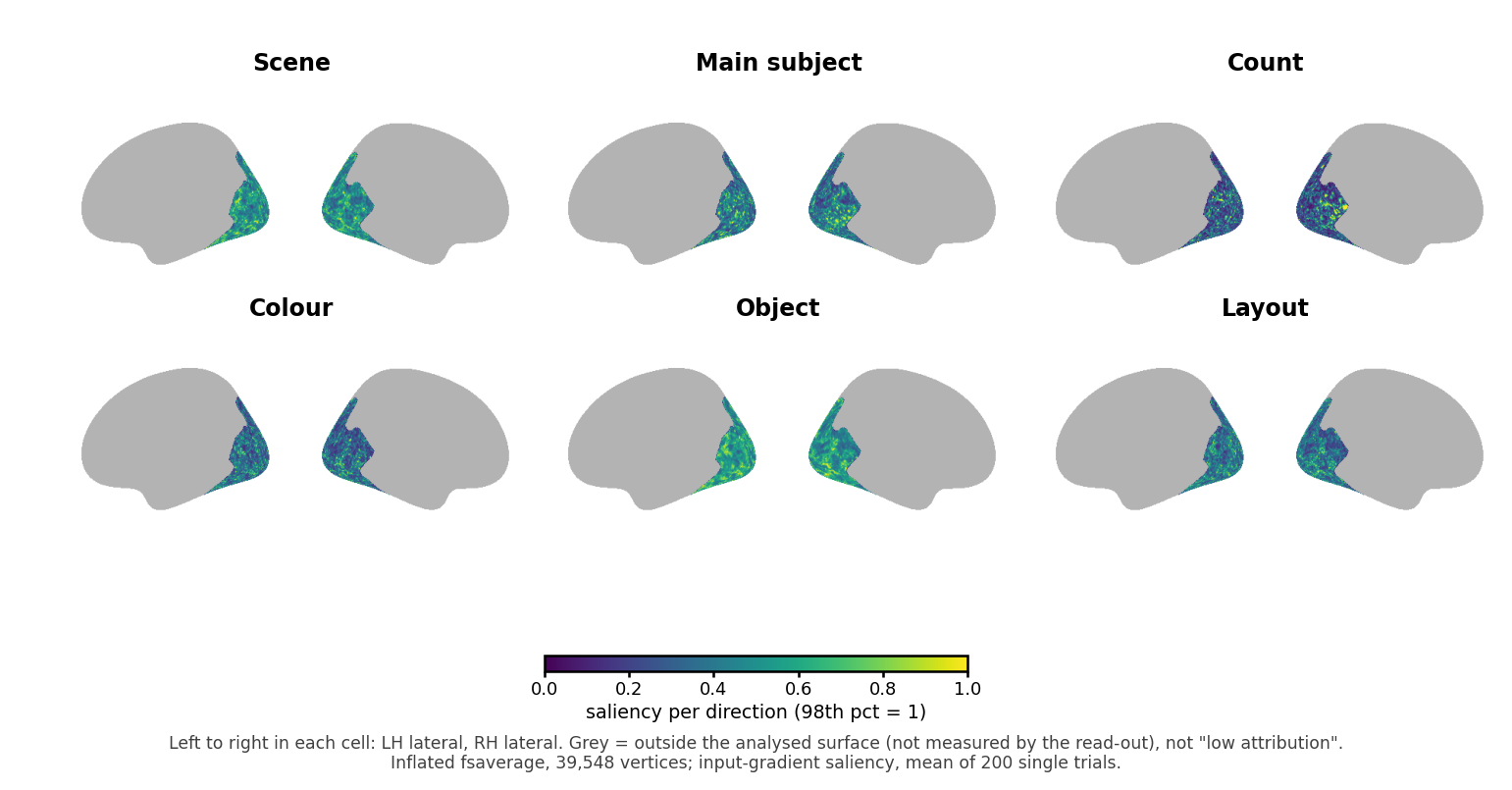}
\vspace*{-0.2cm}
\caption{\textbf{Where each direction reads.} Input-gradient saliency of the six prompt we probed in CLIP read-outs on the inflated fsaverage surface, normalized per head; each cell is LH lateral $\mid$ RH lateral. Every direction loads on the same occipital and ventral-occipital surface; grey marks vertices outside the analyzed surface.}
\label{fig:heads}
\vspace*{-0.2cm}
\end{figure*}

The same holds direction by direction.
\Cref{fig:heads} demonstrates what spatial parts is related to specific prompts. We probe the CLIP readouts with scene,  subject, count, color, object, and layout --- and every one loads on the same occipital and ventral-occipital surface, none of them displaced into a language or memory associated region.
The result differ in \emph{what} they decode, not in \emph{where} they read, which is the descriptive answer to what the content channel sees.
It is also the constraint on that channel: a read-out cannot decode what these vertices do not carry at this signal-to-noise ratio, and \cref{sec:limits} shows what that costs.

\begin{figure*}[t]
\centering
\includegraphics[width=\linewidth]{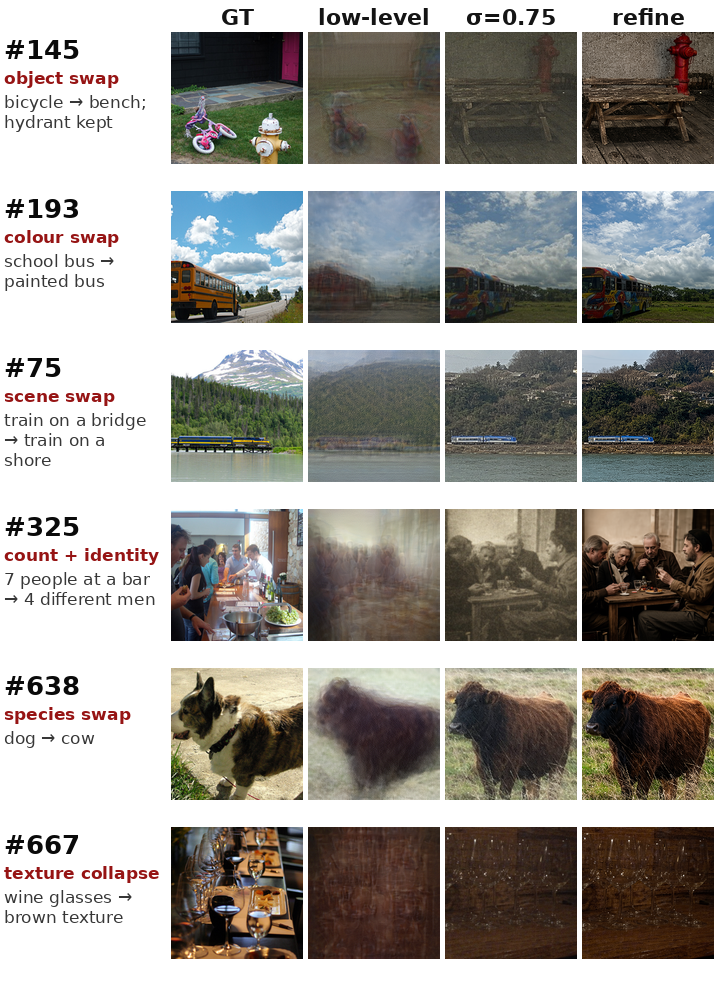}
\vspace*{-0.25cm}
\caption{\textbf{Failure modes}, one row per type, all six taken from the fixed 25-concept draw:
object, color, scene, count-and-identity, species and texture substitution, in the order shown.
Columns: seen image, low-level input, the $\sigma=0.75$ render, NeuroRefine.}
\label{fig:fail}
\vspace*{-0.2cm}
\end{figure*}

\paragraph{How it fails.} Our pipeline is faithful to its input: when the content prompt is wrong, the editor sharpens the mistake rather than repairing it (\cref{fig:fail}).
\emph{Object substitution}: a bird becomes an airplane, a giraffe becomes a zebra.
\emph{Scene substitution}: a fruit basket becomes an indoor scene.
\emph{Count and instance errors}: a crowded race field becomes a single rider, two vases become a room with two people.
\emph{Direction}: facing and mirroring are frequently wrong, because the shared CLIP-like representation is largely invariant to reflection.
These are limits of the content predictor and of the representation, not of the editor; we report them rather than hiding them in an aggregate.
\label{sec:fail}

\subsection{Is the Attribution Causal?}
\label{sec:causal}
\label{sec:anatomy}

\begin{figure}[t]
\centering
\includegraphics[width=0.95\linewidth]{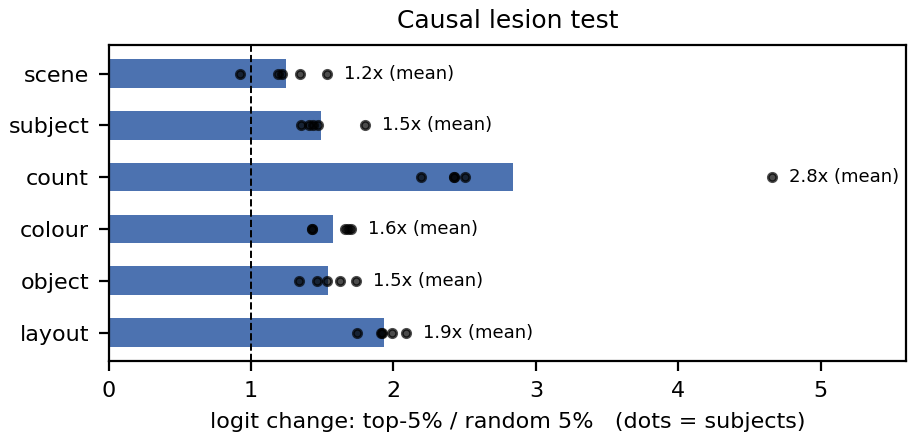}
\vspace*{-0.25cm}
\caption{\textbf{Causal lesion test.} Change in head logits when the top-5\% salient vertices are
zeroed, relative to a random 5\% (dots: individual subjects). Historical ablation: measured on
the retired six-head bank, not yet repeated for the two shipped read-outs.}
\label{fig:lesion}
\vspace*{-0.2cm}
\end{figure}

Descriptively every head reads visual cortex and nothing outside it (\cref{fig:surf,fig:heads}); the question this subsection asks is whether those vertices are load-bearing.
They are: the same attribution, used as a lesion mask, changes the head output several-fold more than a random mask of the same size (\cref{fig:lesion}), and most for the heads that need spatial and numerical structure (count $2.8\times$, layout $1.9\times$, color and object $1.5$--$1.6\times$).

\paragraph{The attribution is causally load-bearing.} Zeroing the most salient vertices changes the head logits several-fold more than zeroing a random set, and more than zeroing the least salient set (\cref{fig:lesion}); the effect is strongest for the heads that need spatial and numerical structure.
Two ceiling measurements frame the interpretation.
The split-half reliability of the repeat trials is near one for the head \emph{outputs}, which only says that the heads are deterministic functions of their input; the reliability of the \emph{fMRI signal itself} per vertex is far lower and is the real ceiling for any readout trained on three repeats (\cref{tab:noise}).
Cross-subject agreement of the attribution maps is small but above a spatial spin null for every head (\cref{tab:crossstat}): there is a shared, weakly-localised substrate, and we do not claim stable individual differences beyond it.

\subsection{What the Pipeline Cannot Fix}
\label{sec:limits}

\paragraph{The editor is not the bottleneck.} Feeding Brain-IT's semantic image to the same frozen editor reaches $0.747$ CLIP cosine and $0.646$ layout, both above what the editor reaches with our content prompt ($0.733$/$0.565$; \cref{tab:ret}).
Since the generator, the step count and the guidance are identical, that gap is the content channel, and no amount of editing recovers it.

\paragraph{Identity and arrangement are a trade.} Retrieval gives the highest identity of any arm ($0.767$) and the lowest layout ($0.207$) because the arrangement it renders belongs to another photograph.
Our arm keeps the arrangement ($0.565$) at an identity cost of $0.034$.
This is why retrieval stays a side comparison: on this panel the two properties move in opposite directions, and only one of them is measured by the CLIP column.

\paragraph{What does not transfer is the read-out.} On BOLD5000 the frozen NSD read-out reaches $56.7\%$ against MindLoop's $84.4\%$, and PixCorr collapses to $0.011$; fitting the read-out on the target subject recovers most of it ($72.5$--$74.3\%$).
The refiner is untouched in all three rows, because it is frozen, so the degradation is entirely in the read-out (\cref{sec:bold}).
The same axis explains the one-hour regime: the read-out trained from scratch stays below the mean-image prior at that budget, and only cross-subject initialisation puts it above the prior from the first session (\cref{sec:onehour}).
A new scanner therefore costs one read-out, not a new pipeline --- and the quality of that read-out is the quality of the method.

\vspace{-0.3cm}

\section{Discussion}
\label{sec:discussion}

\paragraph{What the refiner is and is not.} The refiner is a faithful sharpener, not a second
decoder. The low-level branch supplies a spatial prior, and the prompt retrieved from the predicted CLIP embedding supply the
content; the frozen editor then restores texture and color on top of that layout. When the content
head is wrong, the refiner reproduces the error with high fidelity (\cref{sec:fail}); when the
conditioning is text alone, it re-composes the scene (\cref{fig:ret}). Both observations point to
the same conclusion: the useful information for a refinement stage is the image plus a compact
content signal, and the ceiling of the method is set by the content predictor, not by the low-level
contour.
We measured that ceiling directly. Replacing the text prompt by a caption decoded from the predicted CLIP image embedding (the strongest content signal we have, 0.829 cosine to the true image embedding), and training the decoder conditioning module for that interface, the decoded caption reaches only 0.664 text cosine when conditioned on the true image embedding, which is below the 0.672 of a caption simply retrieved from the pool. The interface, not the brain signal, is the limit, and neither a better low-level contour nor a better selector can lift it.

\paragraph{Sharpness is not quality.} Our outputs are much sharper than Brain-IT's (mean gradient
magnitude $18.5$ versus $9.4$, with the true photograph itself at
$14.9$ under the same $0$--$255$ definition; the earlier pair $0.74/0.27$ was a different scale),
but that number measures edge
energy: over-sharpening and edge
artifacts also raise it, and our own value overshoots the photograph by $24\%$, which we read as a
warning rather than a gain. No LPIPS or human preference study is available here. We therefore use
sharpness only to show that the refiner restores texture the low-level branch lacks, never as a
standalone claim of better perceptual quality.

\paragraph{Why the metrics disagree.} The panel is multi-factor: PixCorr/SSIM/Alex(2) measure
low-level structure, while Inception/CLIP/EffNet/SwAV are semantic. A method can be high on one and
low on the other, and a high PixCorr can coexist with a different content; we quantify the
conflicts in \cref{sec:metric2}. We therefore report the panel factor-wise, with CLIP image cosine
and layout as summaries. We do not claim that the low-level metrics are invalid --- they measure a
different factor --- but a single aggregate hides the trade-off.

\paragraph{Direction.} Facing and mirroring are the one failure that is not inherited from the
content head but from the representation both pipelines share. A CLIP-like embedding is
approximately invariant to left--right reflection, so a direction readout trained on it cannot
recover the sign; a global mirror test corrects only a small fraction of the failures. A dedicated
direction head, trained on explicitly mirrored pairs, and direction-aware evaluation are the
natural next step; we do not claim to have solved direction here.

\paragraph{What transfers, and what does not.} The low-level decoder and the frozen editor are
shared across subjects; only the content readout needs subject data. One hour of it is enough only
with cross-subject initialisation (\cref{sec:onehour}); a single subject trained from scratch at that
budget stays below the trivial mean-image prior, so the ``1\,h'' regime should always be quoted with
that qualifier. Whether the same holds on a different scanner
or dataset is an open question: our external check on BOLD5000 (\cref{sec:bold}) shows that a frozen
NSD readout transfers only partially, and that the refiner amplifies whatever bias the readout
brings.

\section*{Limitations}
(i) The content predictor is the bottleneck: when it is wrong the editor sharpens the error, and
improving it is future work. The refinement stage also trades layout agreement for semantic fidelity,
so we do not claim lossless preservation of the low-level arrangement.
(ii) LPIPS is not computed, following the released Brain-IT code.
(iii) Cross-subject replication covers three subjects.
(iv) The human audit is a curated set, not a random sample.
(v) The low-level metrics (PixCorr/SSIM) are lower for sharper outputs by construction; we report
them as one factor, not as a failure.
(vi) The external BOLD5000 check is preliminary and not directly comparable to the NSD numbers.

\newpage

\appendix
\section*{Appendix}
\section{Additional Qualitative Results}
\label{app:qual}

\begin{figure*}[p]\centering
\includegraphics[width=\linewidth,height=0.94\textheight,keepaspectratio]{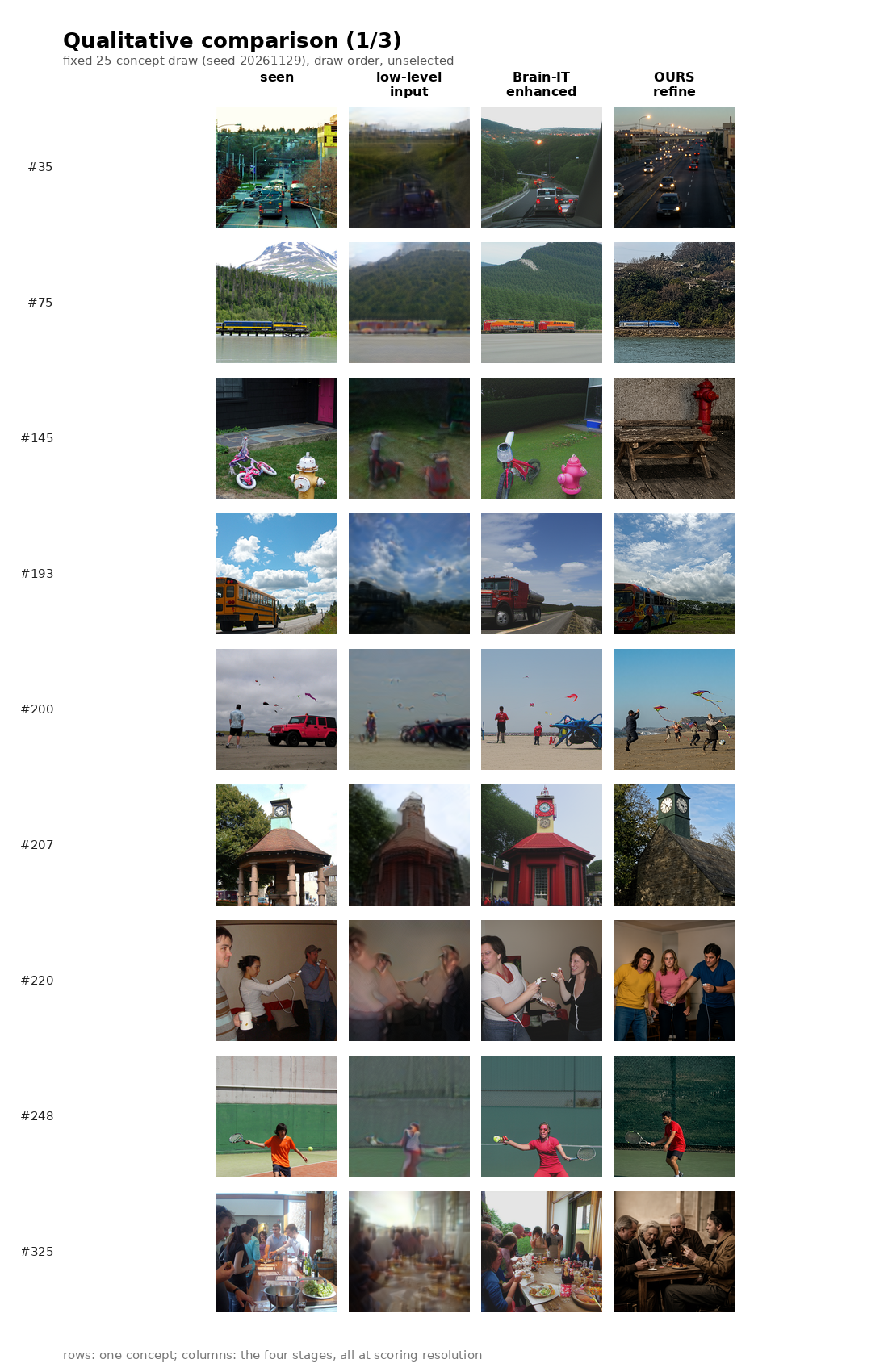}
\caption{Qualitative comparison (1/3): the first 9 of the 25 concepts of the fixed random draw (seed 20261129), in draw order; the figure is unselected. Columns are the seen image, the low-level input, Brain-IT enhanced (reference) and NeuroRefine.}\label{fig:app_c1}\end{figure*}

\begin{figure*}[p]\ContinuedFloat\centering
\includegraphics[width=\linewidth,height=0.94\textheight,keepaspectratio]{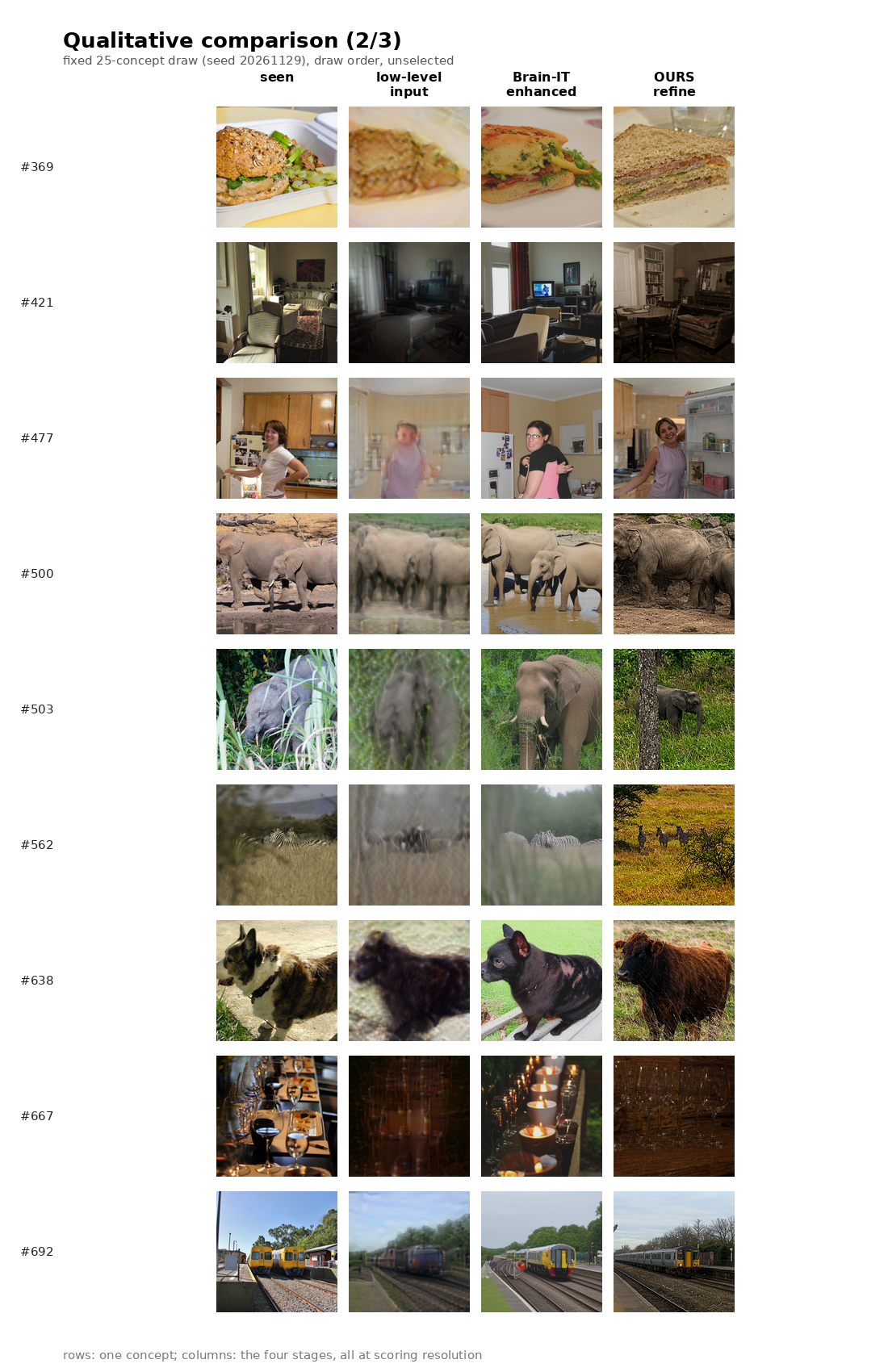}
\caption{(continued) Qualitative comparison (2/3): concepts 10--18 of the same draw.}\label{fig:app_c2}\end{figure*}

\begin{figure*}[p]\ContinuedFloat\centering
\includegraphics[width=\linewidth,height=0.94\textheight,keepaspectratio]{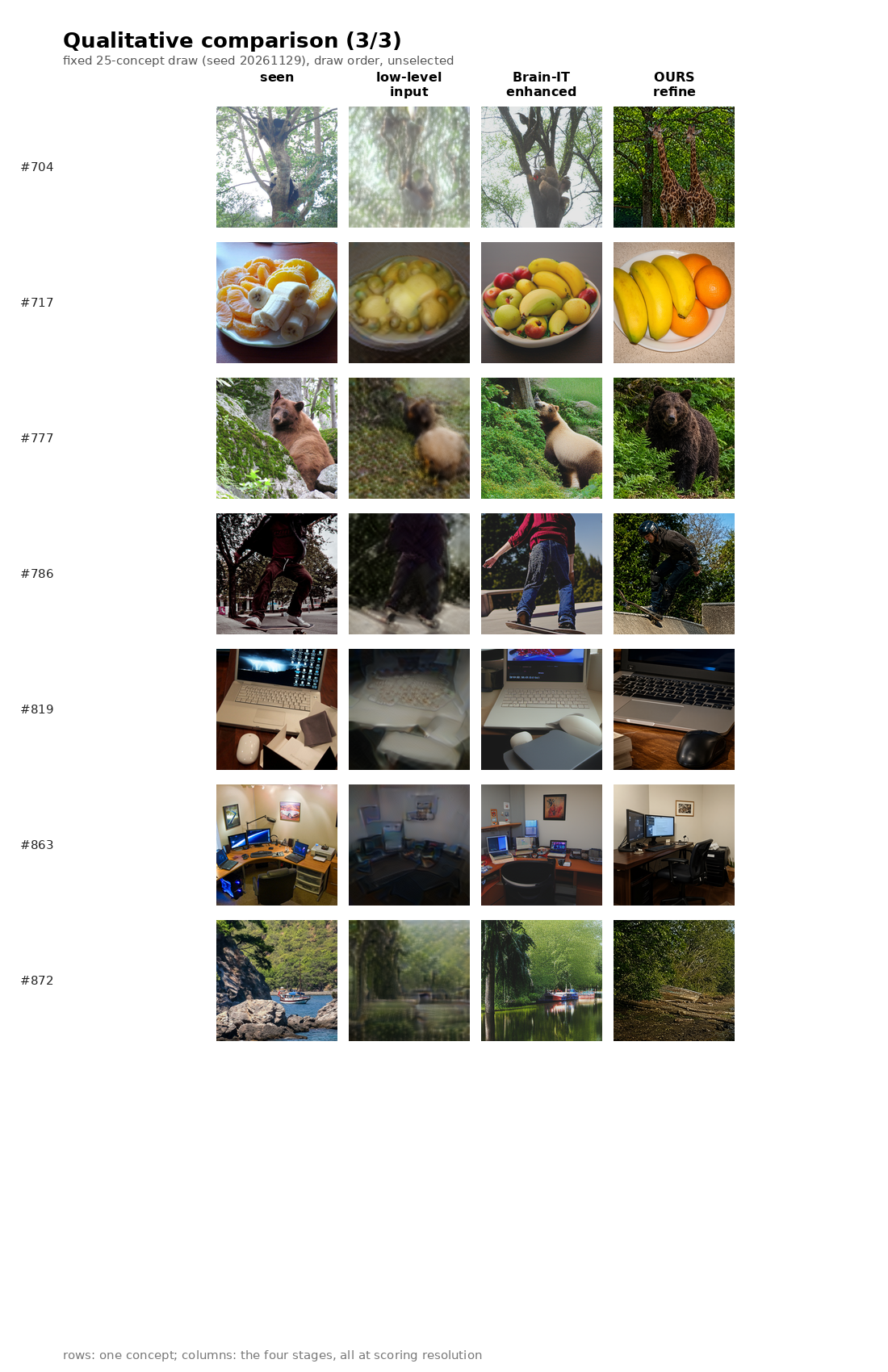}
\caption{(continued) Qualitative comparison (3/3): the remaining 7 concepts of the same draw.}\label{fig:app_c3}\end{figure*}

\begin{figure*}[p]\centering
\includegraphics[width=\linewidth,height=0.94\textheight,keepaspectratio]{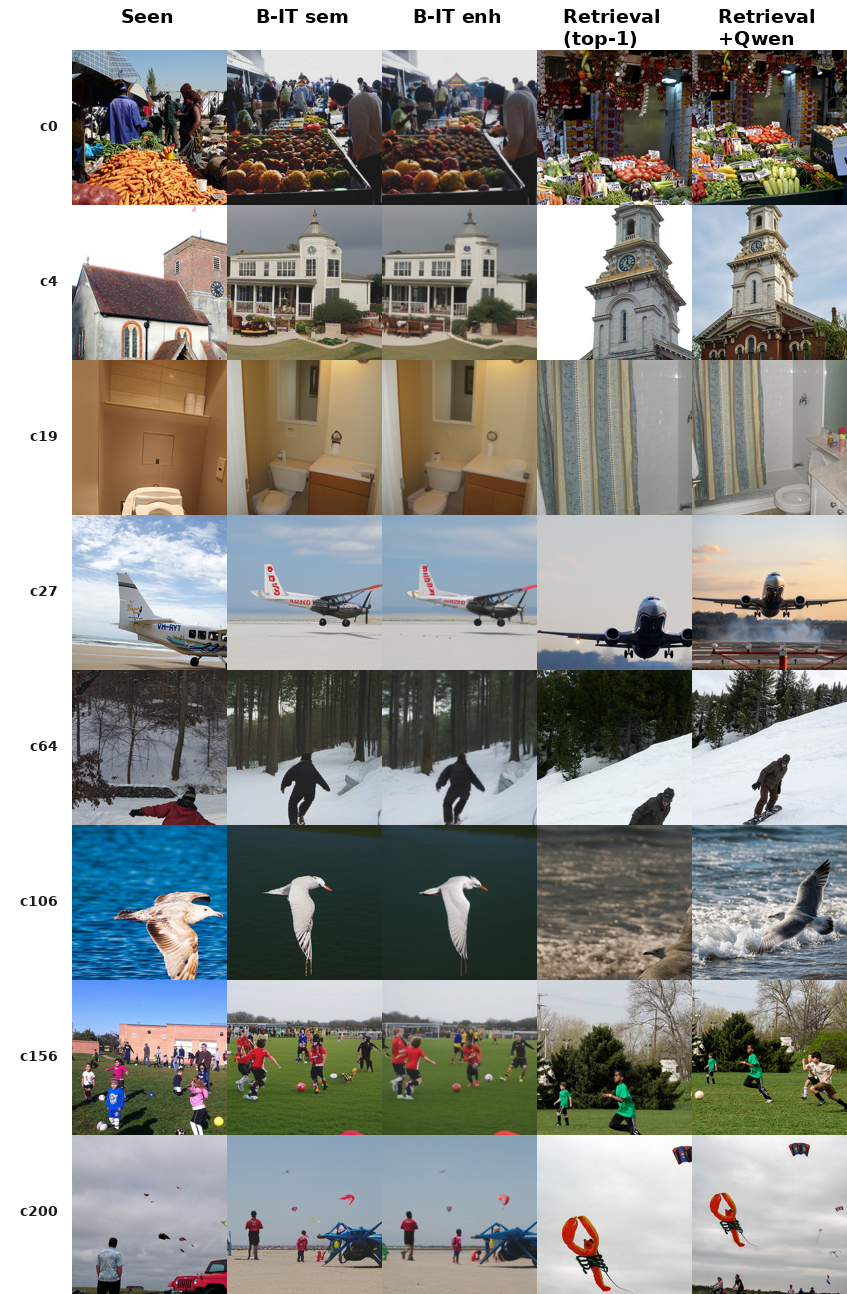}
\caption{Retrieval semantic stage. The retrieved photograph (4th column) is a real image with the
right content but a different arrangement; the edit refiner (last) keeps the retrieved content and
recovers some layout from the low-level input.}\label{fig:app_ret}\end{figure*}

\Cref{fig:app_c1,fig:app_c2,fig:app_c3} continue \cref{fig:cmp} over the 25 concepts of the fixed random draw,
one row per concept and one column per arm, at the resolution used for scoring. They are meant to be
read column-wise: the low-level image supplies a spatial prior and little else; the content prompt
supplies the objects and attributes; and NeuroRefine starts from that arrangement while restoring
texture. Failure modes are visible as well, notably object substitutions inherited from the content
\cref{fig:app_c1} pathway (a dog$\rightarrow$cow in \#638, a bicycle$\rightarrow$bench in \#145), which the editor sharpens rather than corrects.

\section{Additional Quantitative Results}
\label{app:quant}

\begin{table}[h]
\centering\small
\setlength{\tabcolsep}{4pt}
\caption{\textbf{Data efficiency of the semantic readout} on subject 1. ``Transfer'' initialises from
subjects \{2,3,5,7\} and fine-tunes with early stopping. One NSD session is $\approx$250 images
($\approx$1~h of scanning).}
\label{tab:dataeff}
\begin{tabular}{lcccc}
\toprule
Images & from scratch (cos) & transfer (cos) & transfer top-1 & from scratch top-1 \\
\midrule
250 & 0.687 & 0.737 & 0.660 & 0.685 \\
500 & 0.714 & 0.750 & 0.684 & 0.708 \\
1000 & 0.728 & 0.768 & 0.712 & 0.719 \\
2000 & 0.743 & 0.792 & 0.737 & 0.728 \\
8000 & 0.762 & 0.825 & 0.768 & 0.746 \\
\bottomrule
\end{tabular}
\end{table}

The mean-image prior --- the cosine obtained by predicting the dataset-mean CLIP embedding --- is
the trivial baseline for this readout. \Cref{tab:dataeff} therefore says something specific: at the
one-hour budget a readout trained from scratch sits \emph{below} that prior, whereas the same
readout initialised from the other four subjects is already above it, and it stays above at every
budget we tried. The ``1~h'' regime that Brain-IT reports is thus a statement about transfer, not
about single-subject training, and the generator adds no subject data on top because it is frozen.

\section{Head Anatomy: Additional Statistics}
\label{app:anatomy}

\begin{table}[h]
\centering\small
\setlength{\tabcolsep}{4pt}
\caption{\textbf{Repeat-trial reliability of individual vertices} (split-half, Spearman--Brown corrected).
The \emph{fMRI} ceiling is the upper bound for any readout trained on three repeats.}
\label{tab:noise}
\begin{tabular}{lccccc}
\toprule
Subject & 1 & 2 & 3 & 5 & 7 \\
\midrule
split-half $r$ & 0.212 & 0.232 & 0.171 & 0.252 & 0.143 \\
ceiling (mean) & 0.332 & 0.353 & 0.278 & 0.383 & 0.239 \\
ceiling (10--90\%) & 0.10--0.56 & 0.08--0.61 & 0.07--0.49 & 0.14--0.60 & 0.07--0.43 \\
\bottomrule
\end{tabular}
\end{table}

\begin{table}[h]
\centering\small
\setlength{\tabcolsep}{4pt}
\caption{\textbf{Cross-subject agreement of the head attribution maps.} $r_{\text{between}}$ averages all
subject pairs; the spin test rotates one map on the sphere (1{,}000 permutations) and compares the
observed agreement with the null. The non-zero $r$ is small but consistent.}
\label{tab:crossstat}
\begin{tabular}{lcccccc}
\toprule
 & scene & main subj. & count & color & object & layout \\
\midrule
$r_{\text{between}}$ & 0.139 & 0.063 & 0.039 & 0.089 & 0.167 & 0.090 \\
spin null (mean) & $-0.006$ & 0.003 & $-0.005$ & 0.002 & 0.000 & 0.005 \\
spin $p$ & 0.010 & 0.010 & 0.010 & 0.010 & 0.010 & 0.010 \\
\bottomrule
\end{tabular}
\end{table}

Two caveats belong here. First, the within-subject reliability of the head \emph{outputs} is
weight-trivial because the heads are deterministic functions of the betas; it is not evidence of a
reproducible brain map. Second, the between-subject correlation is computed on smoothed surface
maps over visual cortex only, so it quantifies agreement \emph{within} the visual system and says
nothing about the rest of the brain; the spin null is the appropriate test for spatial maps of this
kind, and it is significant for all heads at the finest resolution our permutation budget allows.

\section{Failure Cases}
\label{app:failures}

Our editor sharpens and preserves what the content prompt gives it, so its errors are inherited
rather than invented: when the content predictor confuses objects (a bird becomes an airplane, a
giraffe becomes a zebra) or invents a scene (a fruit basket becomes an indoor scene), the editor
faithfully sharpens the wrong content. Count and instance errors follow the same route: a crowded
race field becomes a single rider, two vases become a room with two people (\cref{fig:fail}).
Direction and object facing are frequently wrong, because both our pipeline and the baselines pass
through a CLIP-like embedding that is largely invariant to mirroring; a global mirror test fixes only
a small fraction of the direction failures, and a dedicated direction readout is future work. These
are limits of the content predictor and of the representation, not of the editor.

\section{Implementation Details}
\label{app:impl}

The pipeline has two frozen stages. Stage~1 is Qwen-Image-2.1 on its distilled four-step schedule at
$512\times512$, with the VAE latent of the low-level image noised to $\sigma=0.75$ as the initial
state, the retrieved content prompt as text, and no image context. Stage~2 is Qwen-Image-Edit-2511 at
$1024\times1024$, $20$ steps, guidance $4.0$, image-conditioned on the stage-1 render and prompted
with the cleanup instruction ``render a clear, high-quality image; fix noise, color noise, exposure
and contrast'' (issued in Chinese).
The frozen generator is shared by every subject and every arm; only the readout is subject-specific. The
retrieval gallery is the union of the NSD images over subjects (66{,}841 images) minus the shared
1{,}000, and we verified that the test set and the gallery are disjoint. All metrics are computed
at $224\times224$ with Brain-IT's \texttt{compute\_metrics}; 2-way identification uses a fixed
random permutation and is stable to $\pm0.002$ across reruns. Sharpness is the mean gradient
magnitude, computed identically for every arm.

\paragraph{Prompt construction across subjects.}
The content prompt is retrieved, not assembled: each subject's predicted CLIP image embedding is
matched against a fixed pool of candidate captions and the nearest one is used, so the prompt differs
across subjects only through the per-subject read-out. The low-level input is subject-independent:
it is Brain-IT's fused low-level image with a fixed unsharp/gamma post-process, used unchanged for
every subject.

\section{Content-Head Label Spaces}
\label{app:labels}

 The six-head bank was trained on labels parsed from Qwen3-VL annotations of the training images
only (retained as an ablation); the 930 shared test concepts are never used for supervision, so no label leaks into evaluation.
The heads read the same full 39{,}548-dimensional vertex vector independently --- they are six
separate one-hidden-layer MLPs, not a transformer and not an attention block, and Brain-IT's
voxel-to-cluster mapping is used only by the low-level branch --- so their only interaction is
through the prompt string that concatenates their outputs. Three fields are single-label and use
categorical cross-entropy on softmax logits (scene, main-subject category, count); three are
multi-label and use independent binary cross-entropy on sigmoid logits (color, object, layout).
The label spaces are:

\begin{itemize}[leftmargin=1.2em,itemsep=1pt,topsep=2pt]
\item \textbf{scene}: 38-way softmax over canonical scenes (market, kitchen, ocean, beach, street,
city, field, forest, mountain, airport, train station, stadium, zoo, \dots) with an \texttt{other}
fallback; the class is assigned by ordered keyword rules on the annotation's scene and caption
fields, most specific rule first.
\item \textbf{main-subject category}: 20-way softmax over the image's main subject (person, people,
man, woman, child, dog, cat, bird, horse, cow, bear, sheep, elephant, giraffe, zebra, fish, airplane,
train, car, food). This is a category of the depicted content, not an NSD participant.
\item \textbf{count}: 3-way (1 / 2 / $\geq$3), taken from the largest number word or digit in the
annotation.
\item \textbf{color}: 12 independent binary labels (white, black, red, blue, green, yellow, brown,
orange, pink, grey, purple, gold). It is multi-label: one image can carry several colors, and each
head output is thresholded independently rather than normalised across classes.
\item \textbf{object}: 150 independent binary labels over a canonical object vocabulary built from
the training annotations. Multiple objects per image are allowed; the head predicts presence, not a
single arg-max class.
\item \textbf{layout}: 8 independent binary labels (left, right, front, behind, above, below,
centre, on), parsed from the relation clauses of the annotation. They describe where the main
subject sits relative to other objects; they are not a dense spatial map, which is why the
low-level image, not this head, carries position.
\end{itemize}

Because color, object and layout are multi-label with independent binary cross-entropy, we report
their accuracy as per-label average precision rather than top-1, and we never compare them
numerically with the single-label scene, main-subject and count heads (which use categorical
cross-entropy).

\section{Metric Definitions}
\label{app:metrics}

\textbf{Cosine similarity}: for embeddings $\bm{a}$ and $\bm{b}$,
\begin{equation}
\cos(\bm{a},\bm{b})=\frac{\langle\bm{a},\bm{b}\rangle}{\lVert\bm{a}\rVert_{2}\,\lVert\bm{b}\rVert_{2}}.
\label{eq:cos}
\end{equation}

\textbf{2-way identification}: for image $i$ and a random other image $\pi(i)$, the trial succeeds if the reconstruction is closer to its own ground truth than to the distractor,
\begin{equation}
\mathrm{Acc}_{2}=\frac{1}{N}\sum_{i=1}^{N}\mathbb{1}\!\left[\cos\big(\bm{E}(\hat{\bm{x}}_{i}),\bm{E}(\bm{x}_{i})\big)>\cos\big(\bm E(\hat{\bm x}_{i}),\bm E(\bm x_{\pi(i)})\big)\right],
\label{eq:twoway}
\end{equation}
and we report accuracy over all images (chance $0.5$).

\textbf{$n$-way identification}: top-1 of $\hat{x}_i$ among the $n$ ground-truth features,
\begin{equation}
\mathrm{Acc}_{n}=\frac{1}{N}\sum_{i=1}^{N}\mathbb{1}\!\left[i=\operatorname*{arg\,max}_{j\in[n]}\cos\big(\bm{E}(\hat{x}_{i}),\bm{E}(x_{j})\big)\right].
\label{eq:nway}
\end{equation}

\textbf{Retrieval recall@$k$}: with the ranking of \cref{eq:retrieval} over the caption pool, and $\mathrm{rank}_{i}$ the position of concept $i$'s reference caption in that ranking,
\begin{equation}
R@k=\frac{1}{N}\sum_{i=1}^{N}\mathbb{1}\!\left[\mathrm{rank}_{i}\le k\right],
\label{eq:recall}
\end{equation}
so $R@1$ is the top-1 of the content-pathway ablations, and $k>1$ measures whether the right caption is merely reachable.

\textbf{CLIP cosine}: cosine between CLIP-L image embeddings of $\hat{x}_i$ and $x_i$,
\begin{equation}
\mathrm{CLIPcos}=\frac{1}{N}\sum_{i=1}^{N}\cos\big(\bm{E}_{\mathrm{img}}(\hat{x}_{i}),\bm{E}_{\mathrm{img}}(x_{i})\big).
\label{eq:clipcos}
\end{equation}
It is one term of the 2-way trial above, not its threshold: the trial also depends on $\cos(\bm{E}(\hat{x}_i),\bm{E}(x_{\pi(i)}))$ for a random distractor, so a model can have a high CLIP cosine and still lose trials when the distractor pool happens to be close.
We report it as the semantic summary because, unlike the 2-way accuracy, it does not saturate near $1$.

\textbf{PixCorr}: Pearson correlation of the two images' pixels,
\begin{equation}
\mathrm{PixCorr}=\frac{\sum_{p}(\hat{x}_{p}-\bar{\hat{x}})(x_{p}-\bar{x})}{\sqrt{\sum_{p}(\hat{x}_{p}-{\bar{\hat{x}}})^{2}}\;\sqrt{\sum_{p}(x_{p}-\bar{x})^{2}}}.
\label{eq:pixcorr}
\end{equation}

\textbf{SSIM}: structural similarity. \textbf{EffNet/SwAV}: $L_2$ distance in the corresponding feature space (lower is better).

\end{document}